\documentclass[twocolumn,aps,prl,showpacs,superscriptaddress,notitlepage]{revtex4-1}
\usepackage{lmodern}

\usepackage[latin9]{inputenc}
\usepackage{amsmath}
\usepackage{amssymb}
\usepackage{graphicx}
\usepackage{epstopdf}
\usepackage{color}
\usepackage{enumerate}
\usepackage{ulem}
\usepackage{natbib}
\usepackage{caption}
\usepackage{soul,color}
\usepackage{comment}

\definecolor{orange}{RGB}{255,125,0}

\begin{document}

	\title{The Critical Current of Disordered Superconductors near T=0}	
		
	\author{A. Doron}
	\email{adam.doron@weizmann.ac.il; Corresponding author}
	\affiliation{Department of Condensed Matter Physics, The Weizmann Institute of Science, Rehovot 7610001, Israel.}
	
	\author{T. Levinson}
	\affiliation{Department of Condensed Matter Physics, The Weizmann Institute of Science, Rehovot 7610001, Israel.}
		
	\author{F. Gorniaczyk}
	\affiliation{Department of Condensed Matter Physics, The Weizmann Institute of Science, Rehovot 7610001, Israel.}	
	
	\author{I. Tamir}
	\affiliation{Department of Condensed Matter Physics, The Weizmann Institute of Science, Rehovot 7610001, Israel.}
	\affiliation{Present Address: Fachbereich Physik, Freie Universit\"{a}t Berlin, 14195 Berlin, Germany.}
		
	\author{D. Shahar}
	\affiliation{Department of Condensed Matter Physics, The Weizmann Institute of Science, Rehovot 7610001, Israel.}
	
	\begin{abstract}
An increasing current through a superconductor can result in a discontinuous increase in the differential resistance at the critical current. This critical current is typically associated either with breaking of Cooper-pairs (de-pairing) or with a collective motion of vortices (de-pinning). In this work we measure superconducting amorphous indium oxide films at low temperatures and high magnetic fields. Using heat-balance considerations we demonstrate that the current-voltage characteristics are well explained by electron overheating that occurs due to the thermal decoupling of the electrons from the host phonons. As a result the electrons overheat to a significantly higher temperature than that of the lattice. By solving the heat-balance equation we are able to accurately predict the critical currents in a variety of experimental conditions. 
The heat-balance approach stems directly from energy conservation. As such it is universal and applies to diverse situations from critical currents in superconductors to climate bi-stabilities that can initiate another ice-age. One disadvantage of the universal nature of this approach is that it is insensitive to the microscopic details of the system, which limits our ability to draw conclusions regarding the initial departure from equilibrium.
	\end{abstract}
	
\maketitle
	
	
\section{Introduction}	
		One of the central properties superconductors is their critical current ($I_{c}$) \cite{larbalestier2011high,kang2006high,hassenzahl2004electric} the maximal current ($I$) they are able to maintain. 	
		The sudden onset of resistance ($R$) at $I_{c}$ is usually associated with one of two mechanisms:  de-pairing, which occurs when the kinetic energy of a Cooper-pair exceeds its binding energy (the superconducting gap) \cite{dew2001critical,tinkham}, or de-pinning of vortices, when the $I$-induced Lorentz-force acting on the vortices exceeds their binding energy, setting them in motion \cite{blatter1994vortices,larkin1975nonlinear,dew2001critical}. Typically, in type-II superconductors under the application of magnetic field ($B$), de-pinning occurs at lower $I$ rendering the de-pairing current a theoretical upper-bound \cite{dew2001critical,tsuei1989limitations}.

Due to the practical significance of $I_{c}$ the bulk of the scientific effort was centered around increasing its value at finite temperatures ($T$'s) rather than on its fundamental, $T=0$, value. In a recent publication $I_{c}$'s of  superconducting amorphous indium oxide films (a:InO) have been studied at low $T$'s and high $B$'s near the high critical $B$ of superconductivity, $B_{c2}$ ($\sim 13$T) \cite{sacepe2018low}.  The authors found that $I_{c}\sim |B-B_{c2}|^\alpha$, with $\alpha \approx 1.6$ that is close to the mean-field value of $3/2$ indicating, as they pointed out, that $I_{c}$ is a result of the combined action of de-pairing and de-pinning where the increasing $I$ initially suppresses the order parameter (by pair-breaking), helping the Lorentz force to overcome the pinning. While by using this approach they were able to suggest a resolution to the linear $B_{c2}(T)$ as $T\to 0$ \cite{tenhover1981upper,hebard1984pair}, their theory is not yet refined enough to offer a quantitative prediction to the value of $I_{c}$ itself. 

Our purpose in this article is to suggest that a different physical mechanism is behind $I_{c}$, which inevitably becomes more dominant as $T\rightarrow 0$. We show that, at high $B$ and very low $T$'s, Joule self-heating induced by the measurement $I$ can result in thermal bi-stability leading to a discontinuous jump in the voltage ($V$) at a well-defined $I$ \cite{gurevich1987self,BStheory,little1959transport,kaganov1957relaxation,wellstood1994hot,kunchur2003hot,kunchur2002unstable,knight2006energy,courtois2008origin,golubkov2000electron,postolova2015nonequilibrium}. Using this approach we are able to predict, within experimental error, the value of $I_{c}$ using only measurements done at $I\rightarrow  0$ and at $I\gg I_c$.

\begin{figure} [h!]
		\includegraphics [height=4.8 cm] {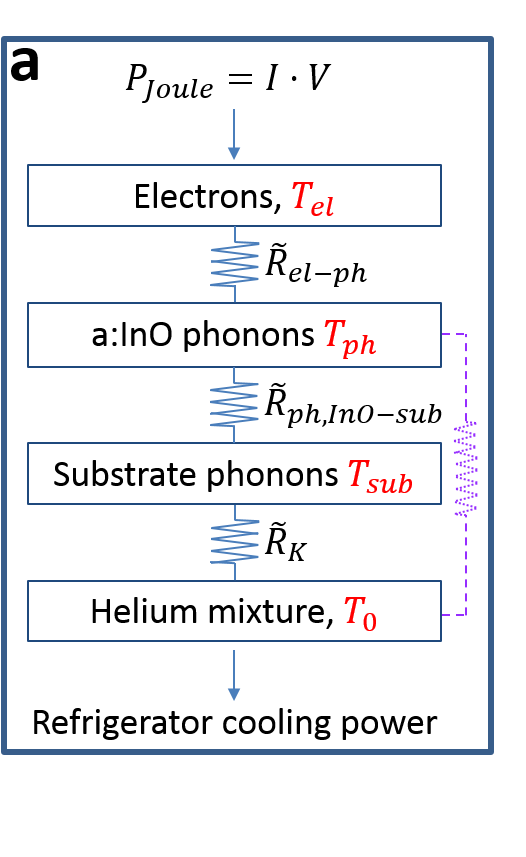}		
		\includegraphics [height=4.8 cm] {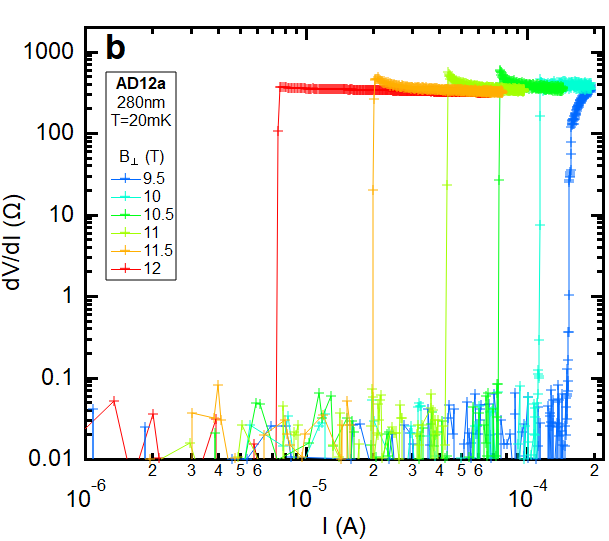}	
		\caption{{\bf } 
(a) Schematic diagram of the heat-flow between subsystems. Joule heating $P_{joule}=I\cdot V$ dissipated in the electrons subsystem by our measurement $I$ flows to the a:InO phonons via a thermal resistor $\tilde{R}_{el-ph}$, then from the a:InO phonons to the substrate phonons via $\tilde{R}_{ph,InO-Sub}$ and into the helium mixture of the dilution refrigerator via $\tilde{R}_{K}$. The purple $\tilde{R}$ stands for other less significant parallel processes $\tilde{R}$'s \cite{FootnoteOtherThermalLinks}. (b) $dV/dI$ vs. $I$ (log-log) at $T=20$ mK of the 280 nm thick film. The color-coding marks different $B_{\perp}$'s where blue is $B_{\perp}=9.5$T and red is $B_{\perp}=12$T. At different $B_{\perp}$'s there are abrupt jumps in $dV/dI$, at  $I_{c}$, which decreases with $B_{\perp}$.
		}			
		\label{Figure1}
	\end{figure}

	Electronic self-heating occurs when the power dissipated by the measurement $I$ exceeds the rate of heat removal from the electrons. To analyze this process we model our experiment as being comprised of four \cite{FootnoteOtherThermalLinks} independent subsystems (Fig.~\ref{Figure1}a) that are thermally coupled via lumped "thermal resistors" ($\tilde{R}$'s):  The electrons, the host a:InO phonons, the substrate phonons and the liquid helium mixture (in which our sample is immersed in our dilution refrigerator). While our system as a whole is driven out of thermal equilibrium by our measurement $I$, it maintains a steady-state where we assume that we can treat each subsystem as being at local equilibrium albeit at different $T$'s represented by $T_{el}$, $T_{ph}$, $T_{sub}$ and $T_{0}$ as indicated in Fig.~\ref{Figure1}a.
	
	Our electronic subsystem is thermally linked to its phonons via $\tilde{R}_{el-ph}$, mediated by electron-phonon coupling \cite{BStheory,little1959transport,kaganov1957relaxation,wellstood1994hot}. The a:InO phonons are, in turn, linked to the substrate's phonons via acoustic transfer at the interface between the different solids \cite{swartz1989thermal}, which transfer their heat to the helium mixture through a thermal-boundary resistance at the interface known as Kapitza resistance, $\tilde{R}_{K}$ \cite{kapitza1941study,johnson1963experiments,pollack1969kapitza}. 
	
Under steady state conditions the power ($P$) flowing across each boundary $\tilde{R}$ is equal to the Joule heating $P\equiv I\cdot V$ delivered to the electronic subsystem. A finite $P$ flowing through the $\tilde{R}$'s results in a $T$-difference between each pair of subsystems. If one of the $\tilde{R}$'s is significantly larger than the others it will constitute a thermal bottleneck, impeding the cooling process, and the largest $T$ difference will develop across it. A straightforward analysis, given in Sec.~S5 of the supplemental material \cite{Supplemental2}, reveals that the thermal bottleneck is between the electrons and the phonons ($\tilde{R}_{el-ph}$) and henceforth we assume that all other subsystems are in equilibrium with each other \cite{FootnoteWLOG}. 
	
The $T$-differences across $\tilde{R}_{el-ph}$ is determined by a heat-balance equation:
	\begin{equation}
		P=\Gamma \Omega (T_{el}^{\beta}-T_{ph}^{\beta})
		\label{eHB}
	\end{equation}
	where $\Omega$ is the sample's volume and $\Gamma$ and $\beta$ are parameters characterizing  $\tilde{R}_{el-ph}$ \cite{BStheory,kaganov1957relaxation,wellstood1994hot}.
	
It turns out that Eq.~\ref{eHB} can lead to dramatic behavior. If a rise in $T_{el}$ that results from an increase in $I$ causes a sufficiently steep increase in $R(T_{el})$, which is certainly the case in our type-II superconductor \cite{FootnoteSteepP}, then below a critical $T_{ph}$ value the heat-balance equation acquires two stable solutions for $T_{el}(I)$: a low $T_{el}$ solution where $T_{el}\gtrsim T_{ph}$ and a high $T_{el}$ solution where $T_{el}\gg T_{ph}$ \cite{BStheory,borisprl}. The jump at $I_{c}$ is simply a manifestation of the system switching discontinuously between the two stable $T_{el}$ solutions, and the sudden increase in $V$ at $I_{c}$ results from $V=I_cR($low $T_{el}\rightarrow$ high $T_{el})$.

We have conducted a systematic study of $I_{c}$ in superconducting a:InO at $0.5>T>0.01$ K and $B$'s $12\ge B\ge 9$ T (where $B_{c2}\simeq 13$T), for samples of various thicknesses in both perpendicular $B$ ($B_{\perp}$) and in-plane $B$ ($B_{||}$). We demonstrate that the $V$-jump at $I_{c}$ results from the behavior expected from the heat-balance Eq.~\ref{eHB}. Furthermore, the value of  $I_{c}$ can be accurately determined using only these considerations. We also show that $I_{c}$ is not consistent with either de-pairing or de-pinning mechanisms, nor with their combined action.  

\section{Experimental} 
	
Our study was performed using 4 a:InO films 26, 57, 100 and 280 nm thick. Each sample was thermally annealed post deposition to a room-$T$ resistivity ($\rho$) \cite{FootnoteEfield} of 4 $\pm 0.2 $ m$ \Omega \cdot$cm, which places them in the relatively low-disorder range of a:InO (see Sec.~S1 of Ref.~\cite{Supplemental2} for additional details regarding the samples). Measurements were performed in an Oxford Instruments Kelvinox dilution refrigerator with a base $T$ of 10 mK, equipped with a z-axis magnet. The samples were mounted on a probe with a rotating head, this allowed us to control the angle between $B$ and the sample plane. While measuring, all conducting lines where filtered using room-$T$ RC filters with a cutoff frequency of 200 KHz. 

\section{Results and analysis}

In Fig.~\ref{Figure1}b we depict several low-$T$ current-voltage characteristics ($I-V$'s) typical of our study. We plot $dV/dI$ vs. $I$ obtained from the 280nm sample measured at $20$mK and at 6 $B$-values between $B=9.5-12$ T ($B_{c2}$ for this sample was $\sim 13$T). All curves exhibit a jump of several orders of magnitudes in $dV/dI$ at a well-defined, and $B$-dependent $I_{c}$: Increasing $B$ towards $B_{c2}$ results in a decrease in $I_{c}$. Similar $B$ dependence of $I_{c}$ is observed in all of our samples.

		\begin{figure} [h!]
		\includegraphics [height=5.5 cm] {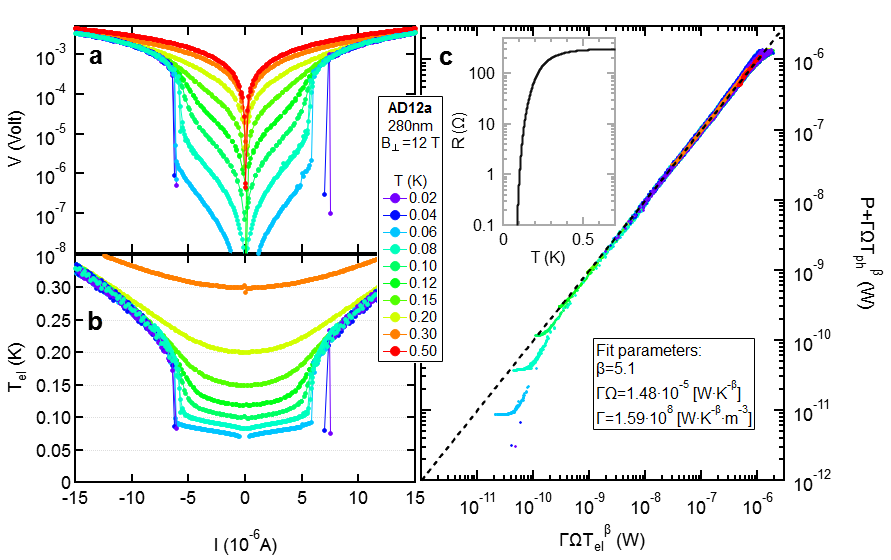}		
		
		\caption{{\bf } 
			(a) $|V|$ vs. $I$ of the 280nm thick sample at $B_{\perp}=12$T. The color-coding marks different $T$'s.
			(b) $T_{el}$ vs. $I$ extracted from the data of (a) using the zero-bias $R(T)$ (inset of (c)) as an electron thermometer.
			(c) Fitting the data to Eq.~\ref{eHB}. By collapsing the different isotherms of (a) such that $P+\Gamma \Omega T_{ph}^{\beta}=\Gamma \Omega T_{el}^{\beta}$ we extract the parameters $\beta$ and $\Gamma \Omega$.
			See Sec.~S3 of the Ref.~\cite{Supplemental2} for additional details. 
		}			
		\label{Figure2}
	\end{figure}

We next demonstrate, using the heat-balance approach (Eq.~\ref{eHB}), that the $I-V$'s are well described in terms of electron self-heating \cite{gurevich1987self,BStheory,little1959transport,kaganov1957relaxation,wellstood1994hot,golubkov2000electron,postolova2015nonequilibrium}. Inspecting Eq.~\ref{eHB} we see that the only unknown variable is $T_{el}$, which we need to obtain independently. For that we assume that all deviations from Ohm's law are due to an increase in $T_{el}$ and not from other non-linear effects (we shall review the flaws of this assumption in the discussion). Under this assumption we convert the raw $I-V$'s obtained from our 280nm film at $B_{\perp}=12$ T at several $T$'s plotted in  Fig.~\ref{Figure2}a to effective $T_{el}$ and plot the results in Fig.~\ref{Figure2}b \cite{golubkov2000electron,maozprl,borisprl} (see Sec.~S3 of Ref.~\cite{Supplemental2} for a detailed description of the heat-balance analysis).

Finally we plot, in Fig.~\ref{Figure2}c, $P+\Gamma \Omega T_{ph}^{\beta}$ vs. $\Gamma \Omega T_{el}^{\beta}$ alongside the fit to Eq.~\ref{eHB} (dashed black line), which our data follow for more than 4 decades, and we extract $\beta=5.1$ and $\Gamma \Omega=1.48\cdot 10^{-5}$W$\cdot K^{-\beta}$. The values of $\beta$ for our samples at various $B$'s are given in table 1 of Ref.~\cite{Supplemental2}. The systematic deviations at low $P$'s are addressed in Sec.~S6 of Ref.~\cite{Supplemental2}. 
	
	\begin{figure}
		\includegraphics [height=5.5 cm] {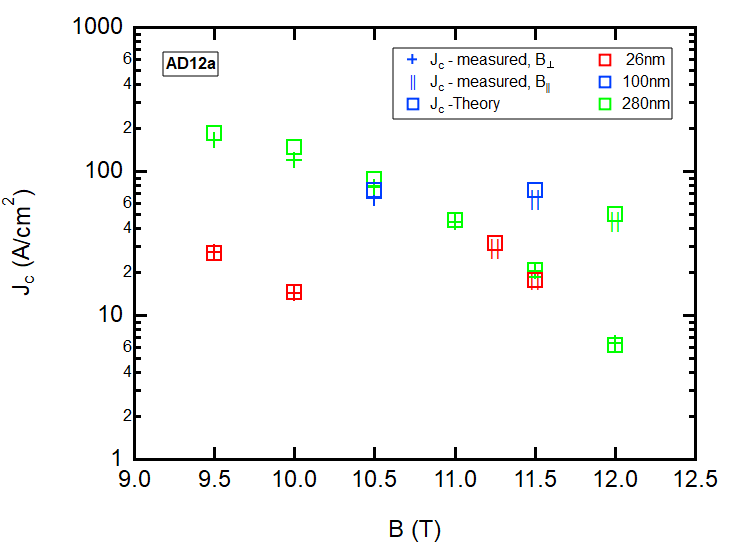}
		\caption{{\bf} 		
			A comparison between measured and theoretical $J_{c}$'s for different samples and $B$ orientations. $J_{c}$ values extracted from the solution of Eq.~\ref{eHB} are marked by squares, crosses and $||$'s mark the measured $J_{c}$ in $B_{\perp}$ and $B_{||}$ respectively. The color-coding marks the sample thickness.
		}			
		\label{Figure3}
	\end{figure}

This leads us to the main result of this work: The broad range of our data that is well described by Eq.~\ref{eHB} enables us to quantitatively predict the values of $I_{c}$ for our superconductors. Using the sample-dependent $\beta$ and $\Gamma \Omega$, together with the measured $R(T)$, we graphically solve Eq.~\ref{eHB} (as detailed in Sec.~S4 of Ref.~\cite{Supplemental2}) and obtain $I_{c}$ for our $B$ and $T$ range. In Fig.~\ref{Figure3} we plot the theoretical 
critical current density $J_{c}$  (squares) for samples with various thicknesses \cite{FootnoteEfield} together with our measured $J_{c}$ (crosses and 
$||$'s representing $B_{\perp}$ and $B_{||}$ respectively). For all samples and $B$ values there is a remarkable quantitative agreement between theory and experiment. We emphasize that the measured value of $J_{c}$ was not used in the heat-balance analysis and so our accurate prediction of $J_{c}$ is a good test for the validity of this theoretical framework.


\section{Discussion}
\subsection{Similarities with insulators}
		
\begin{figure*} [t!]					
	\includegraphics [height=5 cm] {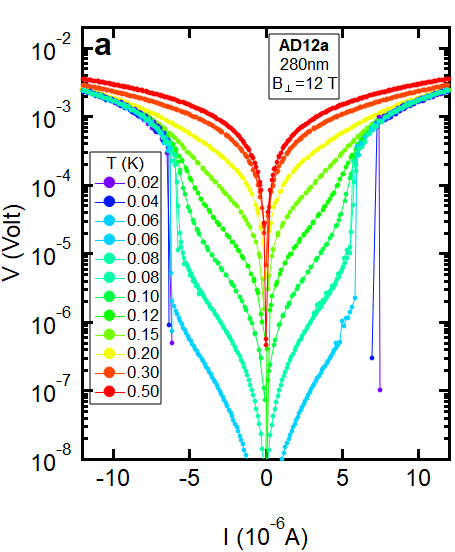}
	\includegraphics [height=5 cm] {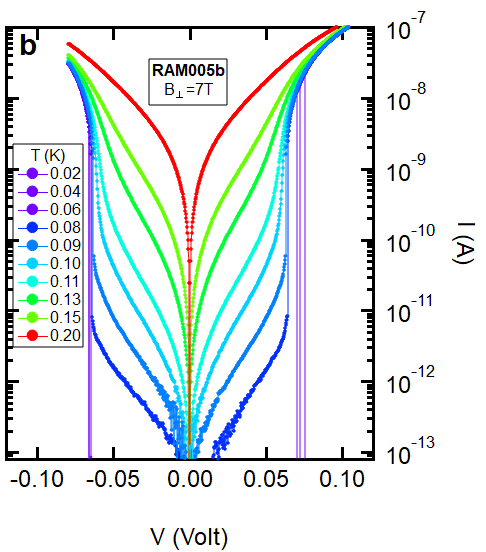}
	\includegraphics [height=5 cm] {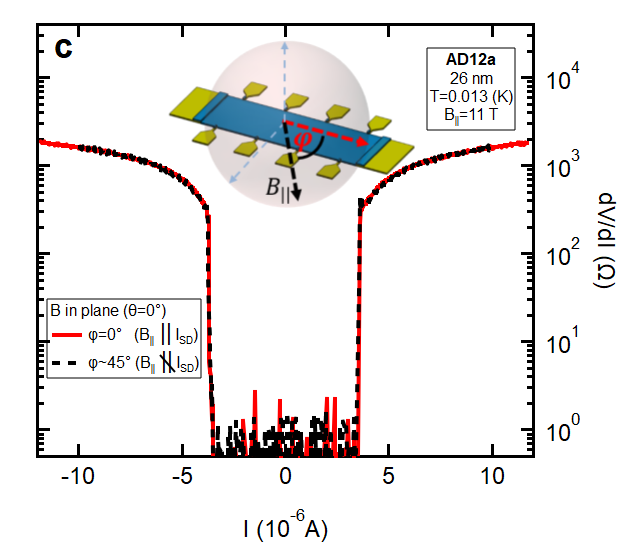}			
	\caption{{\bf } 
		(a) $V$ vs. $I$ obtained from the 280nm thick superconducting sample AD12a at $B=12$T.
		(b) $I$ vs. $V$ obtained from an insulating sample RAM005b at $B=7$T.
		(c) $dV/dI$ (log-scale) vs. $I$ at $T=13$mK and $B_{||}=11$T for the 26nm thick sample at two in-plane angles 
		$\varphi$ (defined in the cartoon) between $B_{||}$ and the source-drain current where the dashed black curve marks $\varphi\sim 45^\circ$ and the red curve marks $\varphi=0^\circ$.
	}			
	\label{Figure4}
\end{figure*}

The heat-balance approach we used throughout this article is a general concept that can account for thermal bi-stabilities in various systems such as superconductors \cite{gurevich1987self,BStheory,kunchur2003hot,kunchur2002unstable,knight2006energy,courtois2008origin}, insulators \cite{borisprl,maozprl}, and even in earth's $T$ \cite{sellers1969global,abbot2018decrease}. Here its use was inspired by earlier studies of the $B$-driven insulating phase of a:InO \cite{murthyprl}. There, the discontinuities in the $I-V$'s were attributed to bi-stable $T_{el}$ assuming that $\tilde{R}_{el-ph}$ dominates the electrons cooling rate at low $T$'s \cite{borisprl,maozprl,levinson2016direct}. 
In Fig.~\ref{Figure4}a and \ref{Figure4}b we plot $V$ vs. $I$ of one of our superconducting samples alongside $I$ vs. $V$ obtained from the $B$-driven insulating phase of a more disordered a:InO sample. The color-coding indicates the measurement $T$. We draw attention to the qualitative similarity between both measurements, and to the fact that the values of the parameters $\beta$ and $\Gamma$ do not vary significantly between superconducting and insulating samples (see table~1 of Ref.~\cite{Supplemental2}).

\subsection{The Ohmic assumption}

In our analysis we assumed that all deviations from Ohmic transport are due to heating and other mechanisms leading to non-linearity, while present, are less effective and do not influence our main results. 
For example, we do not take into account intrinsic non-linearities that are known to exist in type-II superconductors at finite $T$ and $B$ \cite{anderson1962theory,rzchowski1990vortex,van1991dynamics,tinkham}. Our analysis therefore fails to quantitatively account for the onset of non-linearity at $I< I_c$ limiting its range of applicability to $I\ge I_{c}$ and $I\ll I_{C}$. A similar discrepancy was also reported in the heat-balance description of the insulating phase in a:InO \cite{borisprl,maozprl}. We note that self-heating also applies in the presence of intrinsically non-linear effects and a complete description of our $I-V$'s awaits a theory that integrates both self-heating and intrinsic non-linearities.

\subsection{Other mechanisms for $I_{c}$}
The main result of this work is that, at low $T$'s, $I_{c}$ is a result of thermal bi-stability. While we clearly demonstrated this by accurately predicting $I_{c}$ under various conditions, it is also important to show that the other mechanisms for $I_{c}$ are not relevant in our experiments. We focus here on the role played by vortices at a finite $B$ \cite{blatter1994vortices,dew2001critical,larkin1975nonlinear} and refer the reader to Sec.~S2 of Ref.~\cite{Supplemental2} where we  show that the mechanism of de-pairing of Cooper-pairs is unlikely.

To examine whether vortex de-pinning can be the mechanism causing our $I_{c}$'s we oriented $B$ in the sample's plane ($B_{||}$) and conducted two measurements of $I_{c}$: one where $B_{||}$ is aligned parallel to the source-drain $I$ ($I_{SD}$) and one where $B_{||}$ was at an angle of $\varphi \approx 45^{\circ}$ from $I_{SD}$ ($\varphi$ is defined in the inset of Fig.~\ref{Figure4}c).  Because the coherence length of our films $\xi \sim 5$nm \cite{sacepe2015high} is smaller than the film thickness vortices penetrating the plane of the sample experience a Lorentz force $\propto I_{SD}\sin(\varphi)$.
In Fig.~\ref{Figure4}c we plot $dV/dI$ vs. $I$ of the 26 nm thick sample at $T=13$ mK and at $B_{||}=11$T where the dashed black line and the continuous red line correspond to $\varphi \approx 45^{\circ}$ and $\varphi \approx 0^{\circ}$ respectively. 
It is apparent that the entire $dV/dI$ curves, and in particular $ I_{c} $, are completely independent of $\varphi$ demonstrating that $ I_{c} $ is not due to collective de-pinning of vortices \cite{FootnoteHTCvarphi}. 
	
Our $I_{c}$ results are not different from those recently presented in Ref.~\cite{sacepe2018low}. These authors offered an interpretation very different from ours. They claim that $I_{c}$ is a result of a combination of de-pairing and de-pinning. Their main experimental evidence are that $J_{c}\propto |B-B_{c2}|^{\alpha}$ with $\alpha\sim1.6$ which is similar to the mean-field de-pairing value of $3/2$ and that $J_{c}$ is comparable to the de-pairing $J_{c}$ (smaller by a factor of 4 according to their calculation). We do not intend to counter their claims. We do think, on the other hand, that our analysis better describes the data for three reasons: 1. contrary to their results, the value of $\alpha$ is actually non-universal (see Fig.~S2 of Ref.~\cite{Supplemental2}). 2. the de-pairing $J_{c}$ is actually 10-15 times larger than their measured $J_{c}$ and 10-400 times greater than in our measurements (see Sec.~S2 of Ref.~\cite{Supplemental2}) 3. unlike their model the heat-balance analysis provides a good quantitative prediction to $J_{c}$. In the supplemental material of Ref.~\cite{sacepe2018low} Sac{\'e}p{\'e} et al. discuss the possibility of a thermal bi-stability and provide several arguments against this interpretation. In Sec.~S6 of Ref.~\cite{Supplemental2} we respond to these arguments.

\subsection{Lack of hysteresis}		
The heat-balance analysis can only determine the bounds of the $I$-interval where Eq.~\ref{eHB} has two stable solutions \cite{borisprl}. The actual transition occurs stochastically within this interval, depending on a dynamic interplay between the electrons, the phonons and the disorder. By studying the slope of $\log (V)$ vs. $I$ near the discontinuity we can deduce whether the limit of stability is reached (as described in Fig.~2 of Ref.~\cite{borisprl}). In Fig.~\ref{Figure4}a, approaching the high resistive state (HR) to low resistive state (LR) trapping transition, the slope of $\log V$ decreases ("rounds down") prior to the transition, this suggests that the trapping transition occurs at the lower limit of stability. On the other hand, the symmetric "rounding up" of $\log V$ directly after the LR$\to$HR transition suggests that, in our superconductors, the LR$\to$HR transition is triggered prematurely and also occurs near the lower limit of stability, resulting in limited hysteresis.
While the origin of this non-hysteretic behavior remains unclear its explanation awaits further theoretical developments. In Sec.~S7 of Ref.~\cite{Supplemental2} we discuss one possible origin for the limited hysteresis  and present a quantitative comparison between the hysteresis in superconducting and insulating a:InO films.

In summary, we have showed that the $I_{c}$'s of superconducting a:InO films measured at low $T$'s and high $B$'s and are well described by thermal bi-stabilities originating from a model of heat-balance between electrons and phonons (Eq.~\ref{eHB}). 
Using this model we predicted quantitatively $I_{c}$ for samples of different thicknesses for both $B_{\perp}$ and $B_{||}$.
	
	\begin{acknowledgments}
		We are grateful to K. Michaeli, M. V Feigel'man and B. Sac{\'e}p{\'e} for fruitful discussions.
		\paragraph{Funding}
		This research was supported by The Israel Science Foundation (ISF Grant no. 556/17) and the United States - Israel Binational Science Foundation (BSF Grant no. 2012210).
		
	\end{acknowledgments}

	\bibliographystyle{apsrev}

\begin{thebibliography}{46}
\bibitem[{Foo({\natexlab{a}})}]{FootnoteOtherThermalLinks}
	\bibinfo{note}{We neglect other thermal links between subsystems. For example,
		a:InO phonons can transfer heat directly to the liquid helium via Kapitza
		resistance (marked in Fig.~\ref{Figure1}a by a dashed purple line), we omit
		this process as the boundary area between the substrate and liquid helium is
		20 times larger than that of the a:InO and liquid helium}.
\bibitem[{Foo({\natexlab{b}})}]{FootnoteWLOG}
	\bibinfo{note}{The functional form of Eq.~\ref{eHB} is general and also applies
		to the other $\tilde{R}$'s illustrated in Fig.~\ref{Figure1}a, therefore we
		can write the terms of Eq.~\ref{eHB} as if $\tilde{R}_{el-ph}$ is the thermal
		bottleneck and not lose generality.}
\bibitem[{Foo({\natexlab{c}})}]{FootnoteSteepP}
	\bibinfo{note}{It is sufficient that $P$ increases faster than $T_{el}^{\beta}$
		where in our experiment $P=I^{2}R(T_{el})$ and $R(T_{el})\approx
		R_{0}e^{-T_{0}/T_{el}}$}.
\bibitem[{Foo({\natexlab{d}})}]{FootnoteEfield}
	\bibinfo{note}{When comparing samples of different dimensions we use intensive
		parameters such as $J$, $E$ and $\rho$. As our a:InO films are homogeneously
		disordered, they are homogeneous on the relevant length-scales of the sample:
		length ($l$), width ($w$) and thickness ($t$) and we can define $E= V/l$,
		$\rho =Rwt/l$ and $J= I/wt$.}
\end{thebibliography}

\end{document}


\title{Supplemental Material for \\ The Critical Current of Disordered Superconductors near T=0}

\author{A. Doron}
\email{adam.doron@weizmann.ac.il; Corresponding author}
\affiliation{Department of Condensed Matter Physics, The Weizmann Institute of Science, Rehovot 7610001, Israel.}

\author{T. Levinson}
\affiliation{Department of Condensed Matter Physics, The Weizmann Institute of Science, Rehovot 7610001, Israel.}

\author{F. Gorniaczyk}
\affiliation{Department of Condensed Matter Physics, The Weizmann Institute of Science, Rehovot 7610001, Israel.}	

\author{I. Tamir}
\affiliation{Department of Condensed Matter Physics, The Weizmann Institute of Science, Rehovot 7610001, Israel.}
\affiliation{Present Address: Fachbereich Physik, Freie Universit\"{a}t Berlin, 14195 Berlin, Germany.}

\author{D. Shahar}
\affiliation{Department of Condensed Matter Physics, The Weizmann Institute of Science, Rehovot 7610001, Israel.}

\maketitle		

\begin{figure*} 
	\includegraphics [height=4.7 cm] {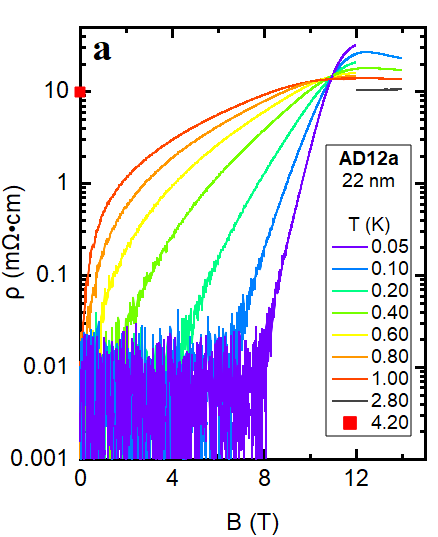}
	\includegraphics [height=4.7 cm] {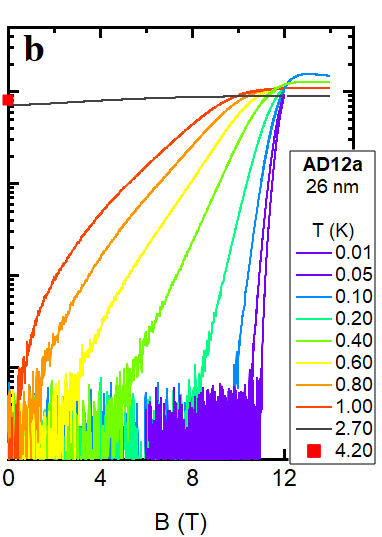}
	\includegraphics [height=4.7 cm] {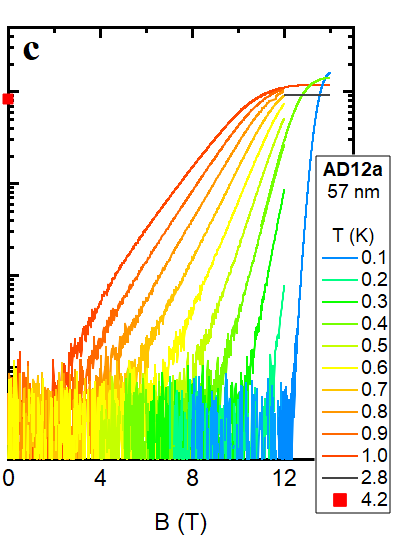}
	\includegraphics [height=4.7 cm] {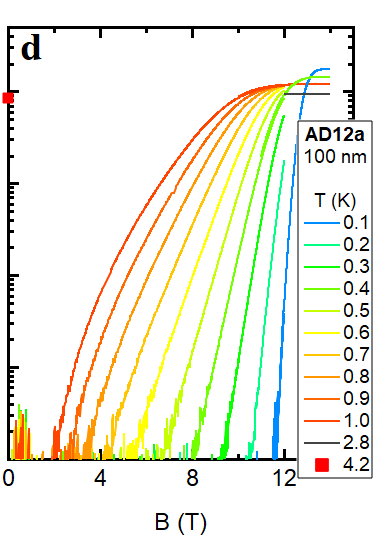}
	\includegraphics [height=4.7 cm] {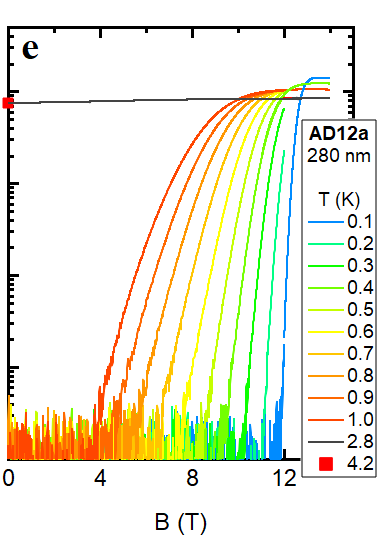}

	\caption{{\bf $\boldsymbol{\rho (B)}$ of low-disordered samples of different thicknesses. } 
		$\rho$ (log scale) vs $B$ of samples of thicknesses (a) 22nm (b) 26nm (c) 57nm (d) 100nm and (e) 280nm.
	}			
	\label{SFigure:RBs}
\end{figure*}

\section{Sample properties and $\boldsymbol{\rho(B)}$}	\label{SamplePrep}

a:InO  was deposited in an Oxygen rich environment of $3\cdot 10^{-5}$ Torr by e-gun evaporation of high purity In$_2$O$_3$ pellets onto a Si/SiO$_2$ substrate (a boron doped silicon wafer with $\rho<5$m$\Omega \cdot$cm with a 580 nm thick oxide layer). 
The sample thickness was measured in situ during evaporation using a crystal monitor and verified later by atomic force microscopy.
The contacts of the samples are Ti/Au, prepared via optical lithography prior to the In$_2$O$_3$ evaporation.
The samples were Hall-bar shaped where the distance between source and drain contacts is 1mm and the width of each sample is 1/3 mm. Adjacent voltage contacts are located 0.8 squares apart (267 $\mu$m). 
In the main text we discuss measurements of three a:InO samples of different thicknesses (26, 100 and 280 nm). The study was actually performed on two more samples of thicknesses 22nm and 57nm. We did not include data of the 57nm thick film in the main-text only because we did not measure sufficiently detailed zero-bias $R(T)$'s of this sample to perform a heat balance analysis. The 22 nm film did not show discontinuities at critical currents, only large non-linearities. We chose to leave the question of why this thinner film did not display a discontinuous response to a future publication.
In order to properly compare between samples, each sample was thermally annealed post deposition to a room $T$ resistivity ($\rho$) of 4 $\pm 0.2$ m$\Omega \cdot$cm, which places them in the relatively low disorder range of a:InO. In figures \ref{SFigure:RBs}a-e we plot $\rho(B_{\perp})$ of each of the five samples where the color-coding marks different $T$'s.

\section{Ruling our de-pairing as the mechanism for $\boldsymbol{J_{c}}$}

\begin{figure} [h!]
	\includegraphics [height=6 cm] {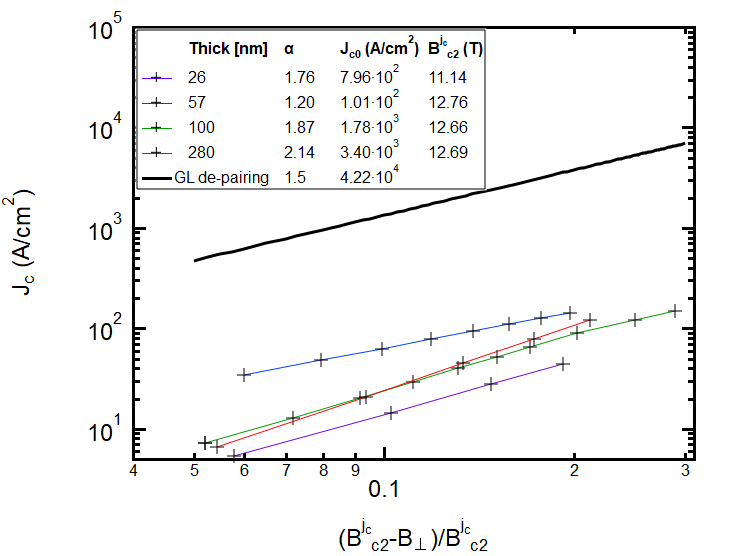}
	\caption{{\bf $\boldsymbol{B}$ dependence of $J_{c}$ - A comparison to the de-pairing $J_{c}$.} 
		$J_{c}$ vs $\frac{B_{c2}^{j_{c}}-B}{B_{c2}^{j_{c}}}$ plotted for four samples of different thicknesses: 280nm (red),  100nm (green), 57nm (blue) and 26 nm thick (purple). $J_{c}$ can be fitted to a power-law according to Eq.~\ref{JcPowerLaw}. The values of the fit parameters for different samples appear in the figure.
		The black line marks the calculated de-pairing $J_{c}$ (according to Eqs.~\ref{eDePairing} and \ref{eGLB0}).
	}			
	\label{SFigure:DePair}
\end{figure}

Raising $I$ through a superconductor increases the kinetic energy of a Cooper-pair. If the kinetic energy exceeds its binding energy (the superconducting gap) Cooper-pairs will break leading to a dissipative state.
This dissipation mechanism is termed the de-pairing mechanism \cite{tinkham}.

The Ginzburg-Landau de-pairing $J_{c}$ in SI units at $T\to 0$ and $B=0$ is \cite{tinkham,dinner2011depairing,MishaPrivate}
\begin{equation}
J_{c0}^{GL}=\frac{\Phi _{0}}{3\sqrt{3}\pi \mu _{0} \lambda ^{2} \xi}
\label{eGLB0}
\end{equation}
where $\Phi_{0}\approx 2.07 \cdot 10^{-15}$ T$\cdot$m$^{2}$ is the magnetic flux quantum, $\lambda$ is the London penetration depth, $\mu_{0}=4\pi \cdot 10^{-7}$ H/m is the vacuum permeability and $\xi$ is the coherence length which is $\approx 5$nm for a:InO samples \cite{sacepe2015high}.
One can estimate $\lambda$ using the relation $\lambda ^{2}=\frac{tL_{k}}{\mu_{0}}=\frac{t\hbar ^2}{\mu_{0}e^2\rho_{s0}}$ where $L_{K}$ kinetic inductance $t$ is the thickness, and $\rho_{s0}$ is the superfluid stiffness at $B=0$.
From Ref. \cite{misra2013measurements} we can extract for a $t=20$nm thick a:InO film $L_{K}\approx3$nH (measured using a two-coil mutual inductance technique). Using Eq.~\ref{eGLB0} results in $J_{c0}^{GL}=4.22\cdot 10^{4}$A/cm$^{2}$.
From Ref. \cite{craneprb751} we can extract for a $t=20$nm thick a:InO film $\rho_{s0}=8\cdot K_{B}$ K where $K_{B}$ is the Boltzman constant (measured using ac conductivity measurements at 9-22GHz). Inserting that in Eq.~\ref{eGLB0} leads to a comparable result $J_{c0}^{GL}=3.23\cdot 10^{4}$A/cm$^{2}$.

From Eq. 5 of Ref. \cite{sacepe2018low} We can calculate the $B$ dependence of $J_{c}^{GL}$ 
\begin{equation}
J_{c}^{GL}(B)= J_{c0}^{GL}(\delta B_{c2})^{3/2}\; ; \; \delta B_{c2}\equiv \frac{B_{c2}-B}{B_{c2}}
\label{eDePairing}
\end{equation} 
The resulting $J_{c}$ vs $\delta B_{c2}$ is plotted as the black line in Fig.~\ref{SFigure:DePair}.

Following the analysis of Ref.~\cite{sacepe2018low}, it turns our that $J_{c}$ of superconducting a:InO films (of a similar disorder level to the films studied in the main-text) also follows a similar power-law behavior which is described in Eq.~\ref{JcPowerLaw}
\begin{equation}
J_{c}(B)=J_{c0}(\delta B_{c2}^{j_{c}})^{\alpha}\; ; \; \delta B_{c2}^{j_{c}}\equiv \frac{B_{c2}^{j_{c}}-B}{B_{c2}^{j_{c}}}
\label{JcPowerLaw}
\end{equation}
where $J_{c0}$ and $\alpha$ and are fit parameters and $B_{c2}^{j_{c}}$ is a set such that  $J_{c}$ will best fit a power-law in $\delta B_{c2}^{j_{c}}$ \cite{FootnoteBc2Jc}. 
In Fig.~\ref{SFigure:DePair} we plot  $J_{c}$ vs $\delta B_{c2}^{j_{c}}$ for four of our films (26nm, 57nm, 100nm and 280nm thick films). 
The values of the fit parameters $J_{c0}$ and $\alpha$ and $B_{c2}$ for each sample are written in the inset of Fig.~\ref{SFigure:DePair}.

Although in both cases $J_{c}$ has a power-law dependence there are significant differences between the measured $J_{c}$ and the calculated de-pairing $J_{c}$: First, the calculated de-pairing $J_{c0}$ is 10-400 times greater than $J_{c0}$ we extract from the fit to Eq.~\ref{JcPowerLaw}. And second, $\alpha$ in the de-pairing description should be $3/2$ \cite{tinkham,sacepe2018low} where we measure a sample dependent $\alpha$ that assumes values between $1.2-2.14$.

One of the key findings in Ref.~\cite{sacepe2018low} is that the measured critical exponent $\alpha$ in three a:InO films is $1.62, 1.65, 1.67 \pm 0.02$ which they note is similar to $3/2$.
Our results show that critical exponent $\alpha$ seems to be sample dependent. For a proper comparison between the findings of the two experiments one should note the following differences:
\begin{enumerate}
	\item We defined $I_{c}$ as the trapping critical current while in Ref.~\cite{sacepe2018low} $I_{c}$ was defined as the "escape" critical current. As the hysteresis is very limited this difference in definition should not be significant.
	\item To properly measure critical exponents one should have a scaling relation that spans over many orders of magnitude in the scaling parameter $\delta B_{c2}^{j_{c}}$. In Fig.~\ref{SFigure:DePair} the scaling is only over a factor of 4-6 in $\delta B_{c2}^{j_{c}}$ and in Ref.~\cite{sacepe2018low} it spans over a slightly larger but still unremarkable factor of 10-20 in $\delta B_{c2}^{j_{c}}$.
	\item In Ref.~\cite{sacepe2018low} $\alpha$ is extracted for samples of a single thickness of 30nm. Here $\alpha$ is extracted for samples of various thicknesses. Note that although our extracted $\alpha$ is not monotonic in the thickness, $\alpha$ of the 26nm thick film is $1.76$ which is not significantly different than the 30nm films of Ref.~\cite{sacepe2018low}.
\end{enumerate}

\section{Detailed Description of the Heat-Balance Fit}	\label{ElectronHeatingAnalysis}

\begin{figure*} [t!]
	\includegraphics [height=4.2 cm] {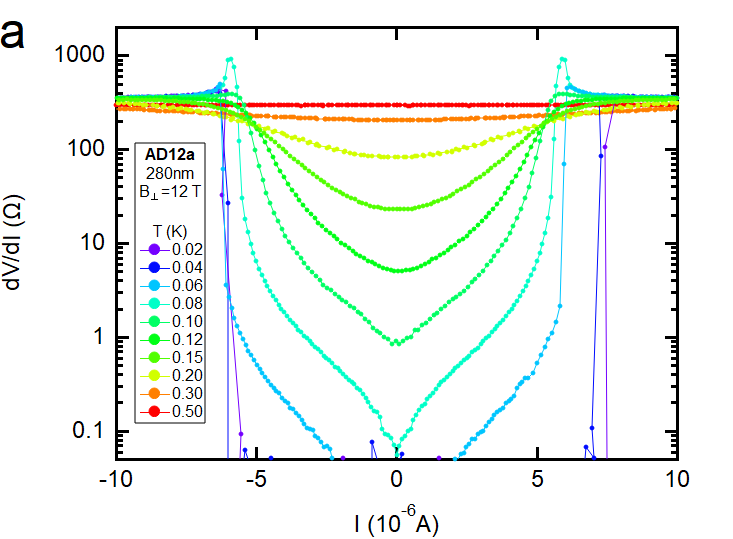}
	\includegraphics [height=4.2 cm] {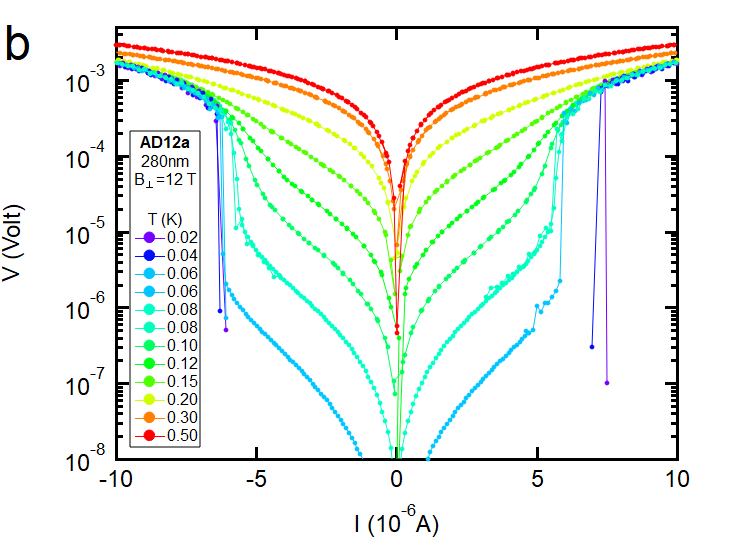}
	\includegraphics [height=4.2 cm] {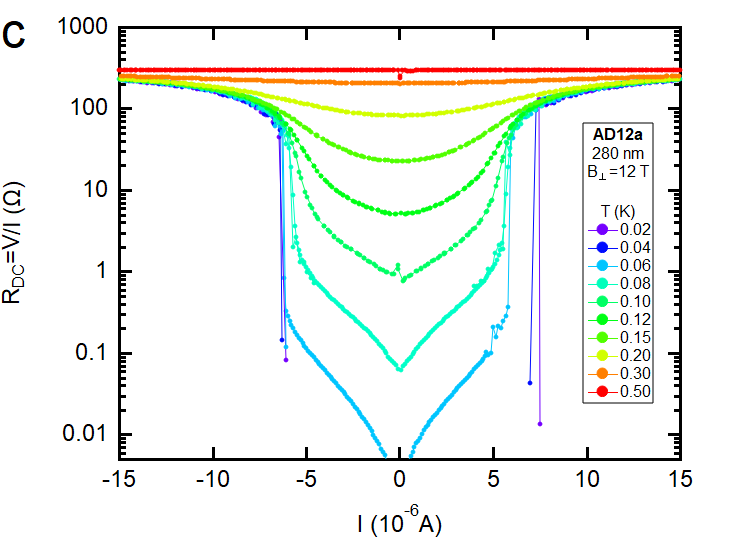}
	\includegraphics [height=4.2 cm] {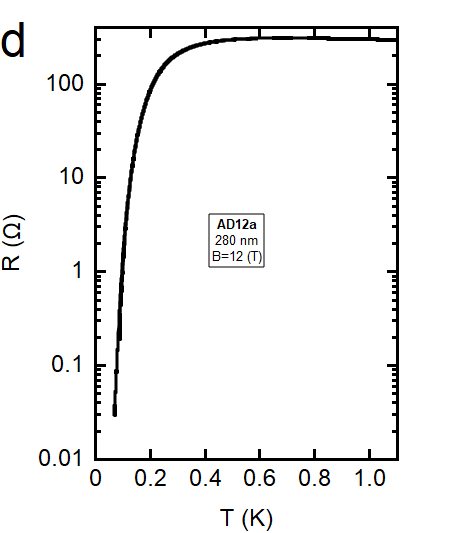}
	\includegraphics [height=4.2 cm] {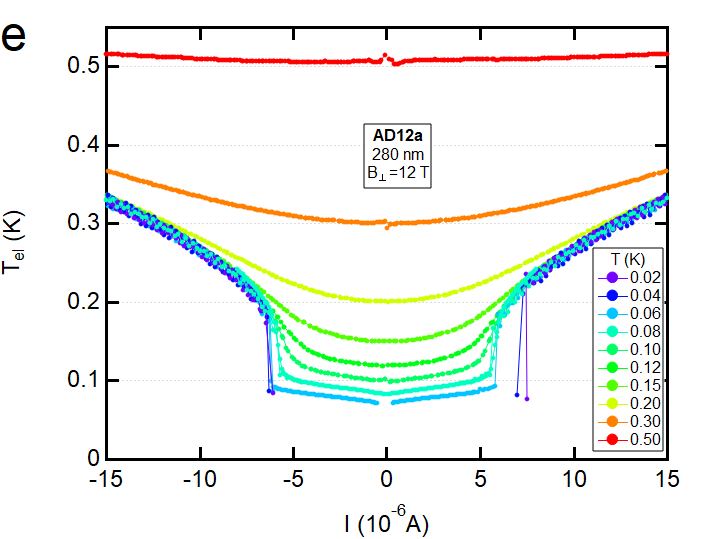}			
	\includegraphics [height=4.2 cm] {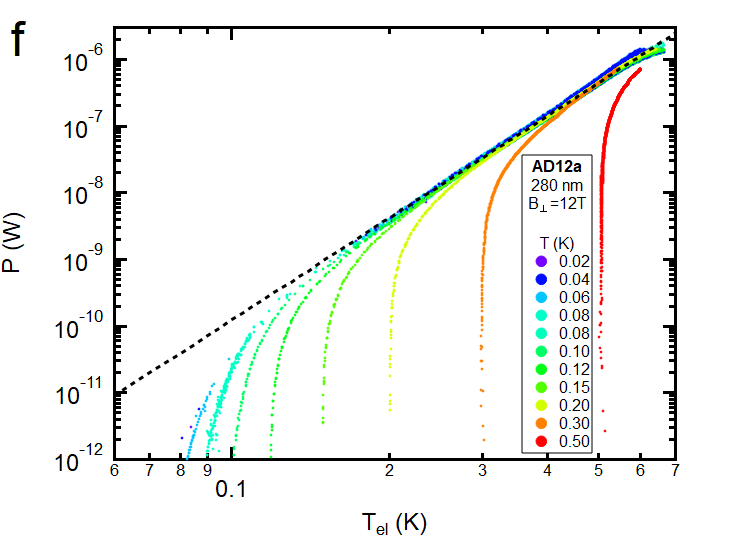}			
	\includegraphics [height=4.2 cm] {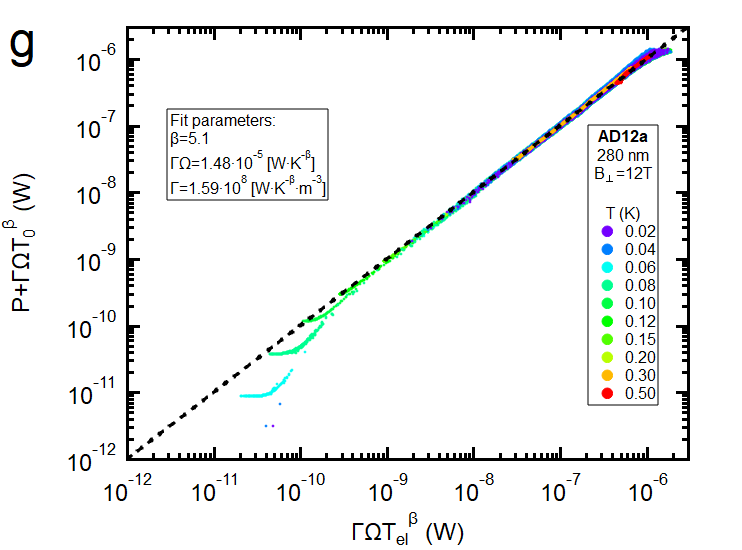}					
	\caption{{Heat-balance analysis} 
		(a) $\frac{dV}{dI}$ vs $I$.
		(b) $V$ vs $I$.
		(c) $R_{dc}\equiv \frac{V}{I}$ vs $I$.
		(d) $R(I=100$nA$)$ vs $T$.
		(e) $T_{el}$ vs $I$.
		(f) $P$ vs $\Gamma \Omega T_{el}^\beta$.
		(g) $P+\Gamma \Omega T_{0}^\beta$ vs $\Gamma \Omega T_{el}^\beta$.		
	}			
	\label{SFigure:HeatBalance}
\end{figure*}

In the main-text we present the results of a heat-balance analysis we performed in order to explain the discontinuities observed at $J_{c}$. 
Here we present a detailed step-by-step account of this analysis.

In Fig.~\ref{SFigure:HeatBalance}a we plot $\frac{dV}{dI}$ vs $I$ of the 280nm thick film at $B_{\perp}=12$T where the color-coding marks different $T$'s.
This measurement was performed in a 4-terminal configuration where in addition to the dc current (x-axis) we also passed a 100nA low frequency (19.19Hz) ac current \cite{FootnoteConstantJinZeroBias}. We simultaneously measured both the resulting ac $V$, from which we plot $\frac{dV}{dI}$ vs $I$ in Fig.~\ref{SFigure:HeatBalance}a and the dc $V$. In Fig.~\ref{SFigure:HeatBalance}b we plot the dc $V$ vs $I$ where $V$ is extracted by a $I$-integration of the ac $\frac{dV}{dI}$ and is consistent with the measured dc $V$.
For the heat-balance analysis we are interested in the dc measurement. The reason for that is two-fold: First, as the dc $I$'s and $V$'s are significantly larger than the ac component (by the design of our measurement) the power dissipated at the sample is $P\sim I_{dc}V_{dc}$. Second, one of the main assumptions of the heat balance analysis is that all non-linear effects in the dc $I-V$'s originate from an elevated $T_{el}$.

The next step is  to extract $T_{el}$ from the $V-I$ data of Fig.~\ref{SFigure:HeatBalance}b \cite{maozprl,golubkov2000electron,postolova2015nonequilibrium}. 
In Fig.~\ref{SFigure:HeatBalance}c we plot $R_{dc}\equiv V/I$ vs $I$ extracted from Fig.~\ref{SFigure:HeatBalance}b. 
In the absence of electron-heating the linearity assumption would result in a constant $R_{dc}$ (as observed in the $T=500$mK red data).
In Fig.~\ref{SFigure:HeatBalance}d we plot a zero-bias $R(T)$ measurement where $I_{dc}=0$ and $I_{ac}=100$nA. As $I_{ac}\ll I_{c}$ and as a reduction of $I_{ac}$ to 10nA did not change the value of $R$ we assume that this measurement was performed in the linear regime therefore we can assume that $T_{el}$ is equal to $T$ of the fridge and use this zero-bias measurement as a calibrated "electron thermometer" for the data of Fig.~\ref{SFigure:HeatBalance}c.
For example $R_{dc}$ of 1 and 10$\Omega$ in Fig.~\ref{SFigure:HeatBalance}c can be translated using the electron thermometry measurement of Fig.~\ref{SFigure:HeatBalance}d to 100 and 130 mK respectively.
The result of this process is plotted in Fig.~\ref{SFigure:HeatBalance}e where we plot $T_{el}$ vs $I$ (data of $T_{el}<90$ mK is achieved by an extrapolation of the $R(T)$ of Fig.~\ref{SFigure:HeatBalance}d and is not significant to any of the conclusions of this work).

In Fig.~\ref{SFigure:HeatBalance}f we plot $P$ vs the resulting $T_{el}$ on a log-log scale where the color-coding marks different $T$'s. 
At sufficiently high $P$ $T_{el}$ is much greater than the fridge's $T$ and all isotherms coincide and follow a power-law. 
The dashed black line is a power-law fit from which we can extract $\beta=5.1$ and $\Gamma=0.595$ nW K$^{-\beta} \mu$m$^{-3}$.
In table \ref{CriticalCurrents:TableBetaAndGamma} we list the values of $\beta$ and $\Gamma$ for several samples at various $B$'s in both superconducting and insulating samples. The dimensions of $\Gamma$ are nW K$^{-\beta} \mu$m$^{-3}$ which depend on $\beta$, therefore, in order to have a proper comparison between samples at different $B$'s and different $\beta$ we multiply $\Gamma$ by $(1K)^{\beta}$.
It is interesting that, although we are comparing samples on both sides of the disorder driven and $B$ driven SIT's, both parameters $\beta$ and $\Gamma$ are always of the same orders of magnitude.
These $\beta$ and $\Gamma$ are used in the main-text to for the graphical solution of the heat-balance equation from which we extract $J_{c}$.

In Fig.~\ref{SFigure:HeatBalance}g we plot the heat balance equation as $P+\Gamma \Omega T_{0}^{\beta}$ vs $\Gamma \Omega T_{el}^{\beta}$ (as was done in Ref.~\cite{maozprl,levinson2016direct,DoronPRL}).
Plotting the data that way we get that all isotherms coincide and data that fits the heat balance equation falls on the dashed black diagonal line.
This fit holds for over 4 orders of magnitude but one can see that there are deviations at low $P$'s. 
There are several possible explanations for the origin of these deviations such as the model being oversimplified and that there is an accumulation of several errors in the translation of $V$ to $T_{el}$ which, at these low $T$'s, becomes comparable to $\Gamma \Omega T_{el}^{\beta}$. We discuss these deviations in Sec.~\ref{DeviationFromHB}.

\begin{table} [h!]
	\centering
	 \begin{tabular}{|c | c | c | c | c|} 
	 \hline
	 Sample name & SC/INS & $B$ [T] & $\beta$ & $\Gamma$~[nW K$^{-\beta} \mu$m$^{-3}\times1K^{\beta}$]\\ [0.5ex] 
	 \hline\hline
	  AD12a 26nm & SC & 9.5$_{\perp}$ & 6.3 & 3.280 \\ 
	 \hline
	  AD12a 26nm & SC & 10$_{\perp}$ & 5.7 & 1.990 \\ 
	 \hline
	  AD12a 26nm & SC & 11.25$_{||}$ & 5.8 & 1.990 \\ 
	 \hline	
	  AD12a 26nm & SC & 11.5$_{||}$ & 5.5 & 1.73 \\ 
	 \hline
	 AD12a 100nm & SC & 10.5$_{\perp}$ & 7.5 & 3.786 \\ 
	 \hline
 	 AD12a 100nm & SC & 11.5$_{||}$ & 7.5 & 3.786 \\ 
 	 \hline
	 AD12a 280nm & SC & 9.5$_{\perp}$ & 9.8 & 4.018 \\ 
	 \hline
 	 AD12a 280nm & SC & 10$_{\perp}$ & 6.6 & 3.053 \\ 
	 \hline
	 AD12a 280nm & SC & 10.5$_{\perp}$ & 6.56 & 1.506 \\ 
	 \hline	 
	 AD12a 280nm & SC & 11$_{\perp}$ & 6.15 & 1.314 \\ 
	 \hline	 
 	 AD12a 280nm & SC & 11.5$_{\perp}$ & 5.6 & 0.948 \\ 
 	 \hline
	 AD12a 280nm & SC & 12$_{\perp}$ & 5.1 & 0.595 \\ 
	 \hline	
	 AD12a 280nm & SC & 12$_{||}$ & 5.96 & 1.456 \\ 
	\hline		 
	 GR12H2a & INS & 3$_{\perp}$ & 8.7 & 2.050 \\ 
	 \hline	
	 RAM005b & INS & 11$_{\perp}$ & 6 & 1.850 \\ 
	 \hline		 	 
	 TL40a & SC & 0.1$_{\perp}$ & 7 & 0.832 \\ 
	 \hline		 
	 TL40a & SC & 0.3$_{\perp}$ & 5.52 & 0.355 \\ 
	 \hline	
	 TL40a & SC & 1$_{\perp}$ & 5.61 & 0.736 \\ 
	 \hline	
	 TL40a & INS & 3$_{\perp}$ & 5.15 & 0.803 \\ 
	 \hline	
	 TL40a & INS & 6$_{\perp}$ & 5.51 & 1.640 \\ 
	 \hline		
	 TL40a & INS & 9$_{\perp}$ & 5.51 & 2.395 \\ 
	 \hline	
	 TL40a & INS & 11$_{\perp}$ & 5.26 & 2.448 \\ 
	 \hline			 	  	 	 	 	 
	\end{tabular}
	\caption{\bf $\boldsymbol{\beta}$ and $\boldsymbol{\Gamma}$ of superconducting and insulating samples}
\label{CriticalCurrents:TableBetaAndGamma}
\end{table}

\section{Graphical Solution of the Heat-Balance Equation}	\label{GraphicalSolution}

The main results of our work is that we can provide a quantitative prediction for $I_{c}$ from the heat-balance analysis.
This prediction of $I_{c}$ is extracted from a graphical solution of Eq.~1 of the main-text which we rewrite below
\begin{equation}
	P=I^{2}R(T_{el})=\Gamma \Omega (T_{el}^{\beta}-T_{ph}^{\beta})
	\label{eHB}
\end{equation}
The measurable quantity that can be calculated from this analysis is $I$ for the transition from the high resistive (HR) to the low resistive (LR) state ($I_{c}^{HL}$) which can be extracted using a graphical solution to Eq.~\ref{eHB}.

\begin{figure*}
	\includegraphics [height=5.4 cm] {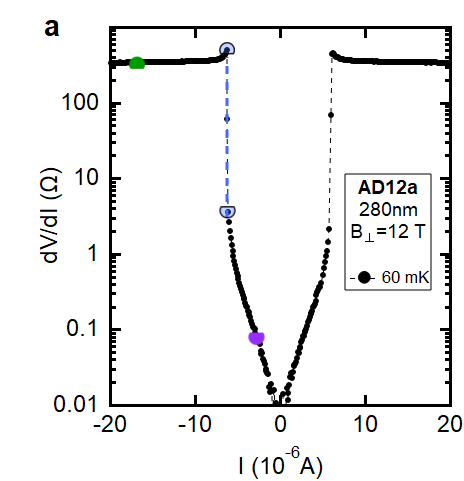}
	\includegraphics [height=5.4 cm] {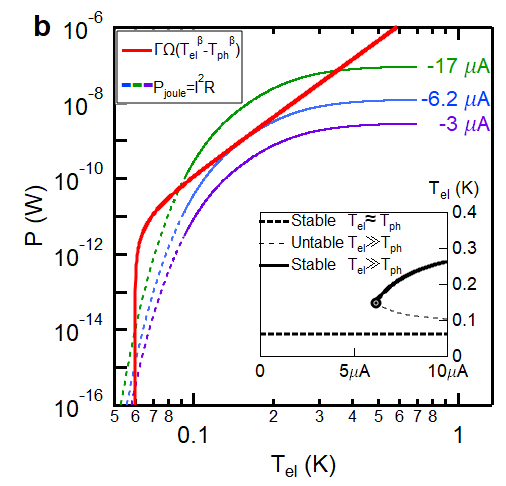}
	\caption{{\bf Graphical solution of the heat-balance equation} 
		(a) $dV/dI$ (log-scale) vs $I$ of the 280nm thick film at $T_{ph}=60$mK and $B_{\perp}=12$T. The purple and green semicircles mark $dV/dI$ at $I=$-3$\mu$A and -17$\mu$A respectively, the two light blue semicircles connected by a dashed blue line mark $dV/dI$ at $I=$-6.2$\mu$A.
		(b) Graphical solution of Eq.~\ref{eHB} for $T_{ph}=60$mK and $B_{\perp}=12$T. The red curve marks the right hand side of Eq.~\ref{eHB} vs $T_{el}$,  the three additional curves correspond to the left hand side of Eq.~\ref{eHB} (joule heating $P=I^{2}R$) where the color-coding (purple, blue and green) corresponds to the same $I$ values of Fig.~(a). A crossing point between the two sides of Eq.~\ref{eHB} occurs at a possible $T_{el}$ solution. The purple, blue and green curves intersect with the red curve once, twice and thrice respectively. 
		Inset: The three possible $T_{el}$ solutions vs $I$. For $I<I_{c}=6.2\mu$A there is a single solution of $T_{el}\approx T_{ph}$ (thick dashed line). For $I>I_{c}$ two more solutions appear, a stable solution (thick continuous line) and an unstable solution (thin dashed line).
		}			
	\label{Figure3}
\end{figure*}

	In Fig.~\ref{Figure3}a we plot the $dV/dI$ vs $I$ for the 280nm thick sample at $T_{ph}=60$mK and $B_{\perp}=12$T. 
	In Fig.~\ref{Figure3}b we plot the graphical solution of Eq.~\ref{eHB}, which accounts for the discontinuities in the data of Fig.~\ref{Figure3}a. 
	The red curve in Fig.~\ref{Figure3}b is the right-hand-side of Eq.~\ref{eHB} where we use the values of $\beta$ and $\Gamma \Omega$ as extracted from the heat-balance analysis. 
	The purple, blue and green curves are the Joule-heating $I^{2}R(T_{el})$ (left-hand-side of Eq.~\ref{eHB}) at $I=-3\mu$A,$-6.2\mu$A and $-17\mu$A respectively and $R$ is the measured zero-bias $R(T)$ (the dashed portion of this line is extracted using a low $T$ extrapolation of $R(T)$ and has no significance to our conclusions). 
	These three $I$ values are also marked in Fig.~\ref{Figure3}a with the same colors.
	A valid $T_{el}$ solution is where each of the three Joule-heating curves intersect with the red curve. In the inset of Fig.~\ref{Figure3}b we plot these possible $T_{el}$ solutions as a function of the driving $I$.

	To illustrate the graphical solution we study each of the three $I$ values separately.
	At $I=-3\mu$A the corresponding purple semicircle in Fig.~\ref{Figure3}a represents a $dV/dI$ in the LR state. 
	In Fig.~\ref{Figure3}b the $I=-3\mu$A purple line intersects with the red curve only once at $T_{el}\approx 0.06$K$=T_{ph}$. This $T_{el}$ solution appears in the inset of Fig.~\ref{Figure3}b as a thick dashed line at $T_{el}\approx 60$mK. 
	In fact for all three $I$'s there is an intersection with the red curve at $T_{el}\approx T_{ph}$ therefore this solution exists for the whole $I$ range plotted in the inset \cite{FootnoteNoTelEqTphSol}.
	At $I=-17\mu$A the corresponding green semicircle in Fig.~\ref{Figure3}a is at the HR state.
	In Fig.~\ref{Figure3}b we see that there are three crossing points between the green and red curves which mark three different $T_{el}$ solutions for Eq.~\ref{eHB}. The middle solution is an unstable fixed point and the low and high $T_{el}$ solutions are stable.
	The unstable $T_{el}$ solution is marked in the inset of Fig.~\ref{Figure3}b by a thin dashed line and the stable high $T_{el}$ solution is marked by a continuous thick line.

	$I=-6.2\mu$A is marked in Fig.~\ref{Figure3}a by a dashed blue line which coincides with $I_{c}^{HL}$ where the sample discontinuously jumps from the HR to the LR state. 
	This value corresponds to a "critical line" in Fig.~\ref{Figure3}b where the blue and red curves are tangent at $T_{el}\approx150$mK, marking the lowest $I$ where a thermal bi-stability can exist.
	At this $I_{c}^{HL}$ the stable and unstable high $T_{el}$ solutions coincide and vanish below $I_{c}^{HL}$ (inset of Fig.~\ref{Figure3}b).

\section{Measurement of the Kapitza resistance}	\label{KapitzaSection}

Eq.~1 of the main-text describes the heat-balance between the electrons and phonons of the a:InO film. 
As mentioned in the main-text, we chose to assume that $\tilde{R}_{el-ph}$ is the largest $\tilde{R}$ and to write Eq.~1 in terms $T_{el}$ and $T_{ph}$ at the outset only for clarity and readability purposes. In fact, as the form of the Eq.~1 is general and describes various heat transfer mechanisms, the heat-balance analysis we performed and the extracted parameters $\beta$, $\Gamma \Omega$ and $T_{el}$ remain valid even if the thermal bottleneck is between two other subsystems.
Below we discuss the scenarios where the thermal bottleneck is between the substrate's phonons and the liquid helium (Kapitza resistance) and between the phonons of the host a:InO of the substrate.

\begin{figure*}
	\centering
	\includegraphics[height=5cm]{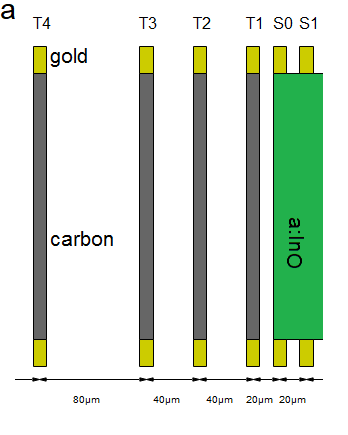}	
	\includegraphics[height=5cm]{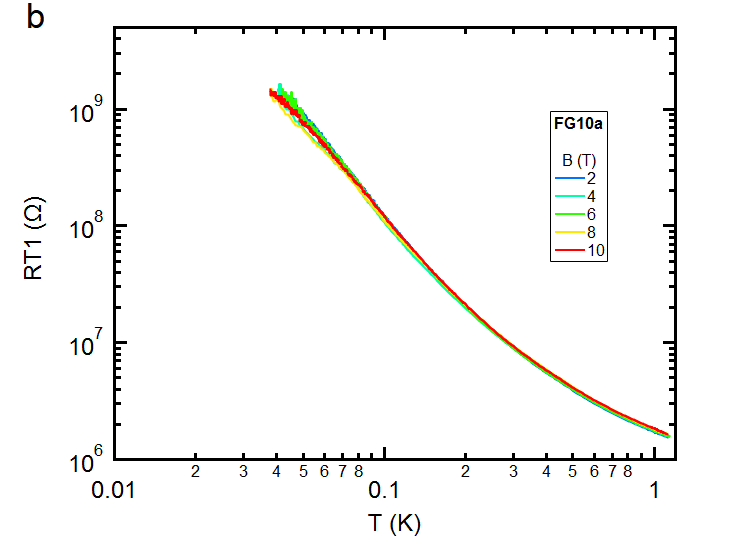}					
	\includegraphics[height=5cm]{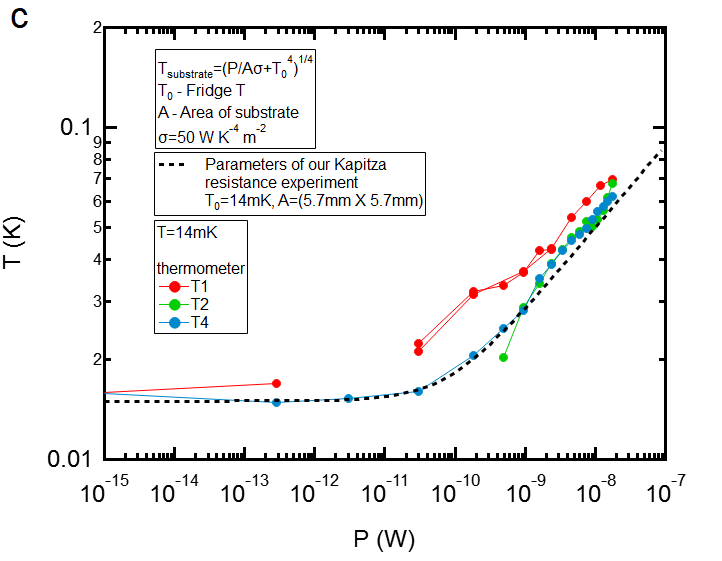}		
	\includegraphics[height=6cm]{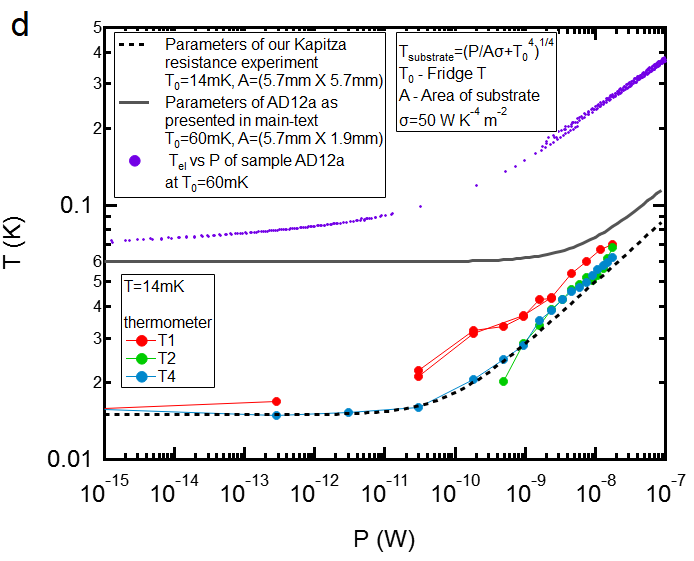}				
	\includegraphics[height=6cm]{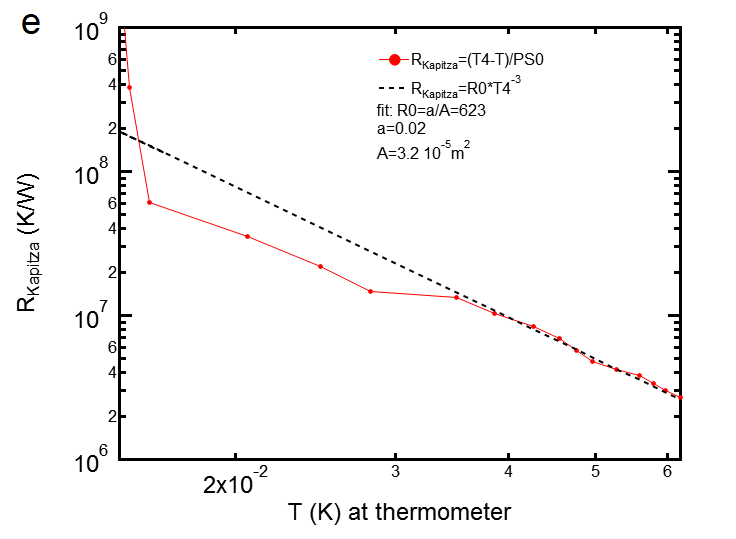}	
	\includegraphics[height=5cm]{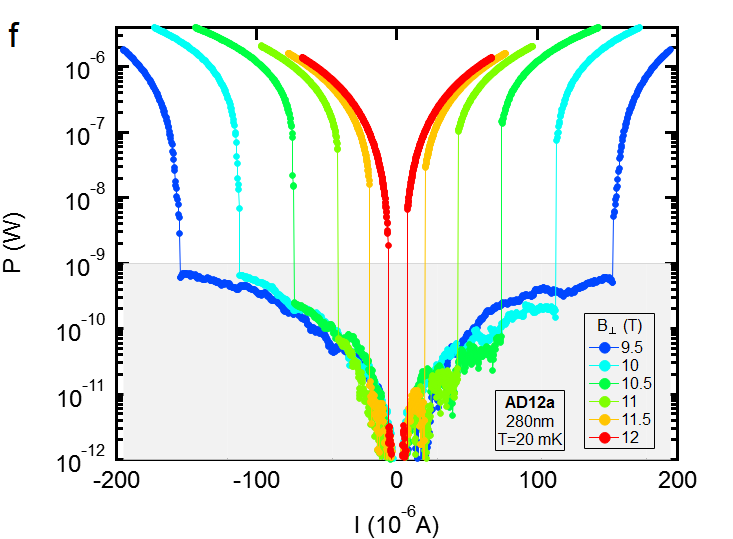}					
	\caption{{\bf Measurement of the Kapitza resistance between the substrate and the helium mixture.} 
		(a) Schematics of the sample. The yellow rectangles mark gold contacts, green marks a:InO (used as a heater) and gray marks carbon paint (used as thermometers). 
		(b) $R$ (log scale) vs $T$ of a carbon paint thermometer. The color-coding mark different $B$'s showing the thermometer is $B$-independent.
		(c) $T$ measured at thermometers T1 (red), T2 (green) and T4 (blue) vs Power dissipated at the heater at $T=14$ mK. The dashed black line is the expected $T_{sub}$ from Eq.~\ref{eTsub} (with no fit parameters) for the parameters of our Kapitza resistance experiment.
		(d) The red, green and blue data points and the dashed black line are the same as in Fig.~(c). The purple data points are $T_{el}$ vs $P$ of the 280nm thick sample at $T=60$mK and $B=12$T (the sample measured in the main-text). The continuous gray line is $T_{sub}$ from Eq.~\ref{eTsub} for the parameters of the sample measured in the main-text.
		(e) $\tilde{R}_{K}$ vs $T$ at thermometer T4 where the red data points are extracted from Fig.~(b) and the dashed black line is a fit achieved using parameters from the literature \cite{pobell2007matter}. 		
		(f) $P$ vs $I$ of the 280nm thick sample at $T=20$mK. The color-coding marks different $B_{\perp}$'s. The data in the gray portion of the figure is within the noise of the measurement. In contrast to what we observe, in the scenarios where the thermal bottleneck is between the substrate and either the liquid helium or the phonons of the a:InO $P$ of the discontinuity should be independent of $B$.
	}			
	\label{KapitzaFig}
\end{figure*}

First is the Kapitza resistance \cite{kapitza1941study,johnson1963experiments,pollack1969kapitza,pobell2007matter}, where cooling is impeded due to an acoustic mismatch between phonons of the liquid helium and of the substrate. If that is the case $T_{el}\approx T_{ph} \approx T_{sub}>T_{0}$.
To test this possibility we conducted an independent experiment and measured the Kapitza resistance of our substrate (a boron doped silicon wafer with $\rho<5$m$\Omega \cdot$cm with 580 nm thick oxide layer).
The schematics of the sample appear in Fig.~\ref{KapitzaFig}a where all patterns where created using optical lithography and a:InO and Ti/Au contacts were prepared as detailed in Sec.~\ref{SamplePrep}. 
Carbon paint thermometers were prepared by first defining their geometry using optical lithography and then immersing the sample in carbon paint until it dries out. A representative $R(T)$ of thermometer T1 at different $B$'s is plotted in Fig.~\ref{KapitzaFig}b showing that the thermometer is insulating and $B$-independent. 
We emphasize that the thermometers are electrically disconnected from the each other and from the heater therefore heat flow is via the substrate.
The thermometers are labeled T1, T2, T3 and T4 (T3 was broken) and the heater is labeled S0 according to the schematics.

In Fig.~\ref{KapitzaFig}c we plot $T$ measured at each thermometer vs the power dissipated at the heater $P$. 
It can be seen that the substrate indeed heats up at sufficiently high powers. 
Following Ref.~\cite{wellstood1994hot} we can estimate the $T_{sub}$ as
\begin{equation}
T_{sub}=(\frac{P}{A\sigma}+T_{0}^{4})^{1/4}
\label{eTsub}
\end{equation}
where $A$ is the area of the substrate (5.7mm $\times$ 5.7mm for the Kapitza resistance experiment and 5.7mm $\times$ 1.9mm for sample AD12a of the main-text), $T_{0}$ is the dilution refrigerator's $T$ and $\sigma=50$ W$\cdot$K$^{-4}\cdot$m$^{-2}$ (see Fig.~9.11 of Ref.~\cite{lounasmaa1974experimental}).
The black dashed line marks $T_{sub}$ calculate from Eq.~\ref{eTsub} for the experimental parameters of our Kapitza resistance experiment. It can be seen that the theoretical description is in excellent agreement with our experimental data measured by thermometers T1, T2 and T4.
We would like to emphasize that we did not use any fit parameters.

In Fig.~\ref{KapitzaFig}d we plot the same data as in Fig.~\ref{KapitzaFig}c and add (purple dots) $T_{el}$ vs $P$ of the 280nm thick sample at $T=60$mK and $B_{\perp}=12$T (the sample measured in the main-text).
The continuous gray line marks $T_{sub}$ calculate from Eq.~\ref{eTsub} for the experimental parameters of sample AD12a at $B_{\perp}=12$T and $T=60$mK. It can be seen that although $T_{sub}$ is elevated it still underestimates $T_{el}$.
For example, at a power of 10nW for the parameters of AD12a at $T_{0}=60$mK, $T_{sub}$ is expected to be 75mK while $T_{el}=247$mK.

For completeness we can extract the Kapitza resistance of our experiment
\begin{equation}
\tilde{R}_{K}\equiv \frac{\Delta T}{P}
\label{eKapitza}
\end{equation}
In Fig.~\ref{KapitzaFig}e we plot $\tilde{R}_{K}$ vs $T$ for thermometer T4.
The dashed black line is $\tilde{R}_{K}=R_{0}T^{-3}$, a functional form used in the literature \cite{pobell2007matter}, where $R_{0}=a/A$, $A$ is the surface area ($5.7\times5.7$mm$^{2}$) and $a=0.02$ as was reported for the thermal boundary of materials with helium mixtures (see section 7.3.3 of Ref.~\cite{pobell2007matter}). 
This functional also has no fit parameters and that it reasonably describes our measured $\tilde{R}_{K}$. 
This suggests again that the phenomenon measured here is indeed increase in $T_{sub}$ due to the $P$ flowing across $\tilde{R}_{K}$.

Another possibility is that the thermal bottleneck is between the phonons of the substrate and of a:InO \cite{swartz1989thermal}. In this scenario $T_{el}\approx T_{ph} > T_{sub}\approx T_{0}$. We did not manage to rule out this possibility but we do view it as unlikely for two reasons; First, the thermal wavelength of the a:InO phonons at $T\approx100$mK in greater than 1$\mu$m therefore larger than the sample's thickness (although not by orders of magnitude). Second, as plotted in Fig.~\ref{KapitzaFig}f, $P$ at the LR side of the discontinuity is highly $B$ dependent where we do not expect the acoustic mismatch between the a:InO and the substrate (and between the substrate and the liquid helium) to have a noticeable $B$ dependence.

\section{Response to the arguments made in Ref.~\cite{sacepe2018low} opposing the bi-stability picture}	\label{DeviationFromHB}
\begin{figure*}
	\centering
	\includegraphics[height=7cm]{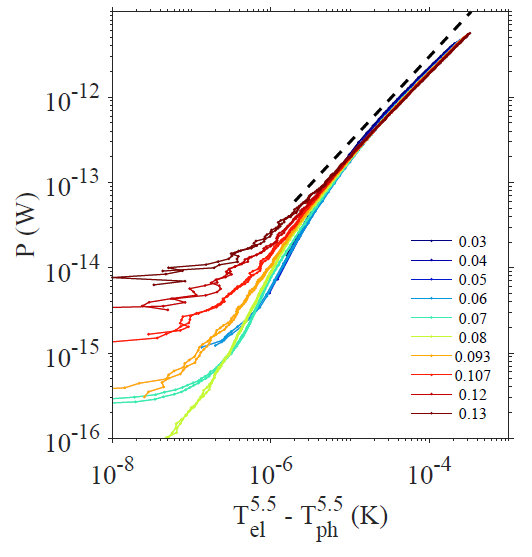}			
	\caption{{\bf Deviation in the Heat-Balance Analysis at Low Power.} 
	$P$ vs $T_{el}^{\beta}-T_{ph}^{\beta}$ with $\beta=5.5$. The figure is takes from Ref.~\cite{sacepe2018low}.
	}			
	\label{CriticalCFigS1}
\end{figure*}

Much of the supplemental material of Ref.~\cite{sacepe2018low} is devoted to explaining why their $I_{c}$'s, measured on a:InO superconducting samples of similar disorder strength to our, are not a result of a thermal bi-stability.
They performed a heat-balance analysis as detailed in Sec.~\ref{ElectronHeatingAnalysis} and in Refs.~\cite{borisprl,maozprl} and plotted $P$ vs $T_{el}^{\beta}-T_{ph}^{\beta}$ as displayed in Fig.~\ref{CriticalCFigS1} which is taken from Ref.~\cite{sacepe2018low}. 
By assuming a power-law dependence, as in Eq.~\ref{eHB}, they extract the parameters $\beta$ and $\Gamma \Omega$. 

Before diving into the details we would like to reiterate that electron-heating theoretical models are simplified models \cite{borisprl,BStheory,kunchur2003hot,kunchur2002unstable,knight2006energy} with some simplified assumptions such as the Ohmic assumption. 
In fact, both in the superconducting data we present in the main-text and in the electron-heating description of the $I-V$'s in the insulating phase there are deviations at low $P$'s (as noted and discussed in Refs.~\cite{maozprl,borisprl}). 
Small deviations from these results are acceptable within the over-heating framework and a full account of these deviations await a theory that incorporates self-heating and intrinsic non-Ohmic behavior. 

One of the main claims of Ref~\cite{sacepe2018low} for rejecting thermal bi-stability as a cause for $I_{c}$ is the scatter in the heat balance analysis at low $P$'s as displayed in Fig.~\ref{CriticalCFigS1} for $P<10^{-14}$W. We claim that, although these deviations can be a result of the over-simplified Ohmic assumption, they are also well within the error of the measurement as they are based on the low $R$ data. This regime is extremely sensitive and any measurement error or noise will be "amplified" by the way the data of Fig.~\ref{CriticalCFigS1} is presented as the x-axis should have very large error bars.


At low $T$'s there will be a non-zero error in $T_{el}$ ($dT_{el}$). This is because the translation to $T_{el}$ is done by comparing $R$ of both the zero-bias measurement and the $V-I$'s (see Sec.~\ref{ElectronHeatingAnalysis}) where at low $T$'s both $R$'s become exponentially small.
At low $I$'s we can assume that $T_{el}\sim T_{ph}$ therefore an error of $dT_{el}$ will translate to $T_{el}^{\beta}-T_{ph}^{\beta}$ (x-axis of of Fig.~\ref{CriticalCFigS1}) as:
\begin{equation}
	\lim\limits _{T_{el}\to T_{ph}}T_{el}^{\beta}-T_{ph}^{\beta}\approx\beta T_{ph}^{\beta-1}dT_{el}
\label{eHBError}
\end{equation}
where we assumed $T_{el}=T_{ph}+dT_{el}$ and $dT_{el}\ll T_{ph},T_{el}$.

Inserting $\beta=5.5$ and approximating $dT_{el}\approx 3$mK \cite{FootnotedTel} for $T_{ph}=70,80,100,130$mK respectively ($\beta$ and $T_{ph}$ of their data) results in error bars in the x-axis of $10^{-7}, 2\cdot 10^{-7},  5\cdot 10^{-7}$ and $1.7\cdot 10^{-6} K^{5.5}$. For all these $T_{ph}$'s such an error bar deems the scatter at low $P$ as insignificant.

A second claim made in the supplemental material of Ref.~\cite{sacepe2018low} is that, comparing to the results of the insulator \cite{maozprl} and the electron-heating theory of the insulator \cite{borisprl},  the $T$ dependence of the LR $\to$ HR  switching current ($I_{escape}$) is too weak, this is because in their heat balance simulation $I_{escape}$ diverges at low $T$'s while their data shows that it does not diverge. 
This claim on their behalf is incorrect on both the theoretical and the experimental levels. 
As discussed in the main-text, the heat balance analysis \cite{borisprl} only predicts the limits of stability, it does not attempt to predict where within this range the jumps will occur (as is stated in the work by Altshuler et al. "As it is usual for the first order phase transition the voltages, where the switches between HR and LR states happen ($V_{HL}$ for HR $\to$ LR and $V_{LH}$ for LR $\to$ HR switches), are determined by kinetics of the decay of metastable states. Theoretical analysis of this decay and evaluation of $V_{HL,LH}$ is beyond the scope of this Letter. Here we can predict only their bounds"). 
Experimentally, in the insulating phase, we typically see that $V_{escape}$ initially increases while cooling (much slower than the maximal $V_{escape}$ predicted by the theory and plotted in the simulation of the supplementary material of Ref.~\cite{sacepe2018low}). At very low $T$'s, not only that it does not increase but it can decrease and saturate at a value similar to $V_{trap}$. This can be seen in the Fig.~3 of Ref.~\cite{doron2017instability} where this phenomenon is discussed in details.
As the self-heating theories \cite{BStheory,borisprl} do not predict the exact values of the transitions it is surprising that the $I_{c}$'s presented in the main-text are predicted quantitatively from the heat-balance equation. 
This point is addressed both in the discussion section of the main-text and in Sec.~\ref{HysteresisSec}.

A third claim  made in the supplemental material of Ref.~\cite{sacepe2018low} is regarding the $B$-dependence of the thermal bi-stability. While calculating the expected HR$\to$LR re-trapping $I$'s from the heat balance model at different $B$'s they got that the re-trapping $I$ should act as a power-law in $|B_{2c}^{j_{c}}-B|^{\alpha}$ with a power of $\alpha=2$. This they write is inconsistent with their measured $\alpha\ge 1.6$.
As discussed above, in Fig.~\ref{SFigure:DePair} we present a similar analysis for several samples and show that the value of this exponent $\alpha$ in non-universal, sample dependent, and can exceed 2.

\begin{figure*}
	\centering
	\includegraphics[height=6cm]{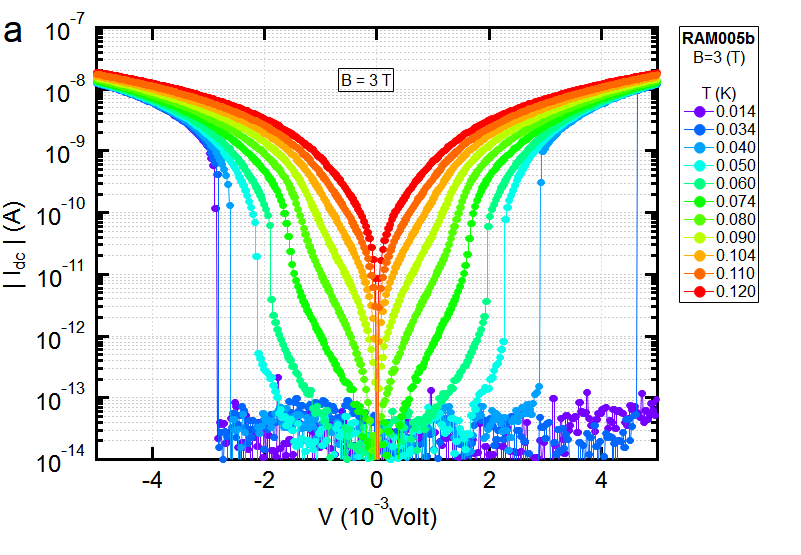}		
	\includegraphics[height=6cm]{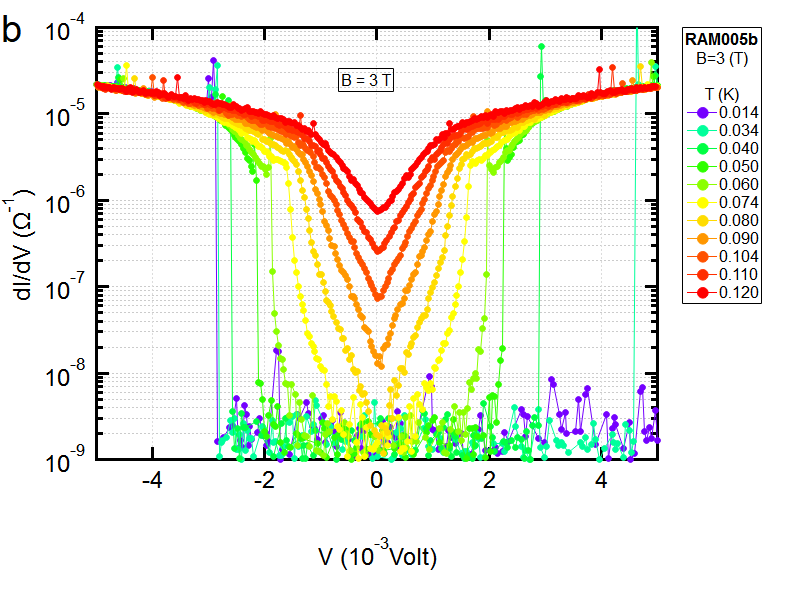}		
	\caption{{\bf} 
		(a) $I$ vs $V$ of sample RAM005b at $B=3$T (in the insulating phase). The color-coding mark different isotherms. (b) $dI/dV$ vs $V$ of sample RAM005b at $B=3$T (in the insulating phase). The color-coding mark different isotherms.
	}			
	\label{RAMB3}
\end{figure*}

A forth claim  made in the supplemental material of Ref.~\cite{sacepe2018low} is that in some of the $B$'s they measure a $T$ dependence in the re-trapping current which in the over-heating picture of the insulator the re-trapping was typically $T$-independent.
As the theory does not predict the exact value of the critical currents it also does not prohibit a $T$-dependence of the re-trapping current. It does predicts that the lower limit of stability can only have a small $T$ dependence. 
Having said that, from our vast experience with electron-heating in the insulating phase the re-trapping voltage is indeed typically $T$ independent. But also in that insulating phase (where the authors of Ref.~\cite{sacepe2018low} write that "The hysteresis and current jump have been proven to be a direct consequence of a thermal bi-stability of the electronic system driven by Joule overheating") this is not always the case as we do sometimes observe $T$-dependent re-trapping currents (as displayed in Fig.~\ref{RAMB3}a).

Their fifth and final claim is their most interesting claim where they point out that their measured $dV/dI$ vs $I$ at low $I$'s is exponential ($\ln (dV/dI)\propto I$), which is consistent with vortex creep below the critical current, while in the heating scenario their simulation shows that $\ln (dV/dI)\propto I^{2}$. 
This is a very interesting claim that we do not have an answer to and we do not see any reason for vortex creep to be absent. 
As discussed in the main-text, such vortex creep is intrinsically non-linear as $\ln(dV/dI)\propto I$ and therefore its expected contribution does seems to be in contrast to one of the central assumptions of the heat-balance model \cite{borisprl} which assumes that any non-linearity is a result of an increase in $T_{el}$.


In the supplemental material of Ref.~\cite{sacepe2018low}  $\ln(dV/dI)\propto I$ vs $\ln (dV/dI)\propto I^{2}$ is used as a "differentiating criteria" between the phenomenon observed in the insulating phase of a:InO and the discontinuities observed in the superconducting phase of a:InO. We do not think that this is a good differentiating criteria. 
For example, in Fig.~S2 of the supplemental material of Ref.~\cite{sacepe2018low} some of the data indeed behaves as $\ln(dV/dI)\propto I$ over some range but a significant portion of their measurements seem to better fit $\ln (dV/dI)\propto I^{2}$ (see Fig.~S2 of their supplementary material, $B=10.5$ at $T>120$mK, $B=11$T at $T>100$mK, $B=11.15$T at $T>80$mK, $B=11.4$T at $T>60$mK and $B=11.5$T at $T>50$mK). We note that the $\propto I^{2}$ in their data is mostly at $T$'s where the jump begins to diminish but these $T$'s are still much smaller than the typical activation $T$ which they relate to thermally assisted flux-flow (for example, in Fig~S3 they show that at $B=11.25$T the activation $T$ is $0.75$K where in Fig.~S2 at $B=11.4$T and $T\ge 69$mK $\ln(dV/dI)\propto I^{2}$ ).
This is consistent with our superconducting films where in both $B_{\perp}$ and $B_{||}$ we observe that $\ln (dV/dI)$ sometimes better fits $\propto I^{2}$ than $\propto I$.
On the other hand in Fig.~\ref{RAMB3}b we plot $dV/dI$ vs $V$ measured on sample RAM005b at $B=3$T in the insulating phase where we see that although they claim that such discontinuities are due to electron heating, $\ln(dV/dI)\propto V$. 
This shows that the $\ln(dV/dI)\propto I$ vs $\ln (dV/dI)\propto I^{2}$ criteria is not lacking.

We summarize that, as the Ohmic assumption is merely an approximation, some deviations at low $P$'s are acceptable within the over-heating framework. Such deviations at low $P$'s in the insulating phase of a:InO are presented and discussed in Refs.~\cite{maozprl,borisprl}. Some of the claims made in Ref.~\cite{sacepe2018low} against the electron-heating model are focused on this low $P$ regime. A full account of these deviations awaits a theory that combines self-heating and intrinsic non-Ohmic behavior.

\section{Lack of Hysteresis}	\label{HysteresisSec}
\begin{figure*}
	\centering
	\includegraphics[height=6cm]{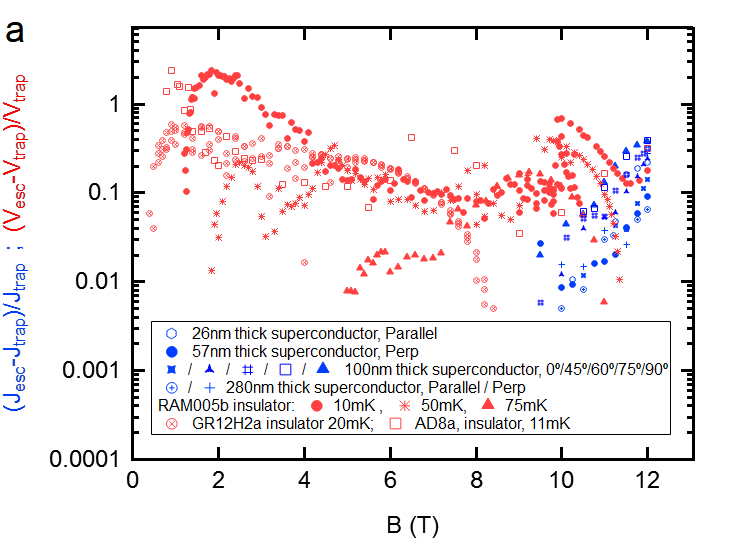}	
	\includegraphics[height=6cm]{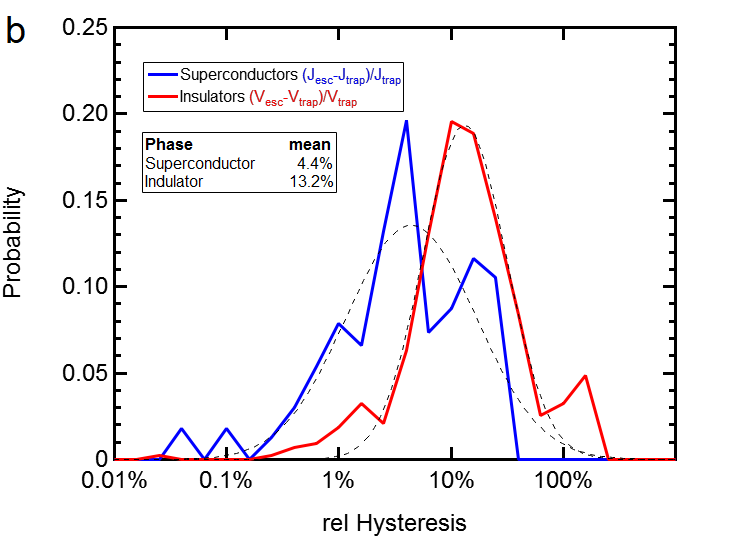}		
	\caption{{\bf Relative hysteresis in the superconducting and insulating phases of a:InO.} 
		(a) Relative hysteresis vs $B$. The blue data corresponds to $\frac{J_{esc}-J_{trap}}{J_{trap}}$ of the superconducting samples discussed in the main text. The red data marks $\frac{V_{esc}-V_{trap}}{V_{trap}}$ for a:InO samples in the insulating phase. The hysteresis was measured at various $T$'s and $B$ orientations as noted in the inset. 
		(b) The distribution of relative hysteresis extracted from the data of (a) where the blue and red mark the recurrence of the relative hysteresis in the superconducting and insulating phases respectively. The dashed black lines mark log-normal fits to the data. The mean value of the relative hysteresis is noted in the inset.
	}			
	\label{RelativeHysteresis}
\end{figure*}

In the main-text we noted that the LR$\to$HR discontinuity is triggered prematurely, resulting in a limited hysteresis.
So far we were unable to fully account for this observation. Below we present a quantitative analysis of the hysteresis and compare it to the hysteresis measured in the insulating phase of a:InO.

We define the relative hysteresis between $J_{c}$ in the escape and trapping sides of the transition in the superconducting phase as $\delta J\equiv \frac{J_{esc}-J_{trap}}{J_{trap}}$ and the  relative hysteresis between $V_{c}$ in the escape and trapping sides of the transition in insulating samples as $\delta V \equiv \frac{V_{esc}-V_{trap}}{V_{trap}}$.
In Fig.~\ref{RelativeHysteresis}a we plot $\delta J$ for superconducting samples in blue and $\delta V$ for insulating samples in red. The relative hysteresis includes different $B$'s, $T$'s, and $B$ orientations.
In Fig.~\ref{RelativeHysteresis}b we plot the distribution of relative hysteresis extracted from the data of Fig.~\ref{RelativeHysteresis}a for superconducting (blue) and insulating (red) samples. 
As the distribution of relative hysteresis is spread almost normally over several orders of magnitude we chose the bin-sizes in Fig.~\ref{RelativeHysteresis}b to be logarithmically spaced (bin sizes of equal $\log{(\delta J)}$ and $\log{(\delta V)}$).
The mean relative hysteresis (using a log-normal distribution) for the insulating samples we investigated is 13.2$\%$ and for superconducting samples it is 4.4$\%$.
While the relative hysteresis in the insulating phase is indeed three times larger than in the superconducting phase, for the time being we are unable to draw any conclusions from this difference.

The premature triggering of the escape transition is also observed in insulating a:InO samples and was previously interpreted as a result of the high disorder in the samples \cite{maozprl,borisprl,KravtsovPrivate,doron2017instability}. As this is a possible explanation for the premature triggering of the LR$\to$HR transition we repeat this explanation below.
The discontinuous jumps are considered to be akin to 1st order phase-transitions such as the Van-der-Waals liquid-gas phase transition \cite{DoronPRL,borisprl} where the transition does not occur at the limit of stability but according to the Maxwell area law. One can push the transition towards the limits of stability by adding nucleation centers. The role of nucleation centers in our sample can be taken by hot spots that might form locally near defects in the sample. As a result, both LR$\to$HR and HR$\to$LR transitions can be triggered prematurely, at the lowest end of the bi-stability interval.

\bibliographystyle{apsrev}